%% file: paperTR.tex
\documentclass[conference]{IEEEtran}
\usepackage{epsfig,amsmath,subfigure,rotating,moreverb,pst-tree}

\newcommand{\ignore}[1]{}

\newcommand{\eat}[1]{}
\newcommand{\name}{$\operatorname{RTChoke}$}
\newcommand{\rednote}[1]{\color{red}{ \em #1 }\color{black}}
\newcommand{\bluenote}[1]{\color{blue}{ \em #1 }\color{black}}
\renewcommand{\rednote}[1]{}
\renewcommand{\bluenote}[1]{}

\def\unit#1#2{\hbox{#1$\,\textrm{#2}$}}
\newcommand{\rateEntry}{r_{\operatorname{in}}}
\newcommand{\rateExit}{r_{\operatorname{out}}}
\newcommand{\currSamplingInterval}{S}
\newcommand{\numSamplesAtTime}{n}

\newcommand{\numSamplesReqd}{N}
\newcommand{\numCars}{c}
\newcommand{\timeReq}{T}
\newcommand{\mymax}{\operatorname{max}}

\begin{document}

%don't want date printed
\date{}

%make title bold and 14 pt font (Latex default is non-bold, 16 pt)

\title{RTChoke: Efficient Real-Time Traffic Chokepoint Detection and Monitoring}

\author{\IEEEauthorblockN{Vikram Munishwar}
\IEEEauthorblockA{State University of New York, Binghamton.\\
vmunish1@cs.binghamton.edu}
\and
\IEEEauthorblockN{Vinay Kolar, Praveen Jayachandran, Ravi Kokku}
\IEEEauthorblockA{IBM Research, India\\
\{vinkolar,praveen.j,ravkokku\}@in.ibm.com}
}

\IEEEoverridecommandlockouts
\IEEEpubid{\makebox[\columnwidth]{978-1-4244-8953-4/11/\$26.00~\copyright~2015 IEEE \hfill} \hspace{\columnsep}\makebox[\columnwidth]{ }}

\maketitle

%\vspace{-60pt}

% Use the following at camera-ready time to suppress page numbers.
% Comment it out when you first submit the paper for review.
%\thispagestyle{empty}
%\eatreminders{}

\input{abstract}

%\vspace{-0.1in}
%\vspace{-0.1in}
\input{introduction}
%\vspace{-0.05in}
\input{motivation}

\input{related}

%\vspace{-0.1in}
\input{designTR}
%\input{implementation}
%\vspace{-0.05in}
\input{analysisTR}
%\vspace{-0.05in}
\input{experiments_blr}
\input{experiments_sfoTR}

%\vspace{0.05in}
\input{conclusion}

%\vspace{-0.1in}

\begin{small}
\bibliographystyle{IEEEtran}
\bibliography{paper}
\end{small}

\end{document}

%% file: abstract.tex
\begin{abstract}

We present a novel efficient adaptive sensing and 
monitoring solution for a system of mobile sensing devices that support traffic monitoring applications. 
We make a key observation that much of
the variance in commute times arises at a few congestion {\em hotspots}, and a reliable estimate
of congestion can be obtained by selectively monitoring congestion just at these hotspots.
%While the high  up-front cost of sensor infrastructure makes it prohibitive for governments to invest in 
%them, the proliferation of smart phones has made several mobile sensors readily available.
We design a smartphone application and a back-end system that automatically identifies and monitors
congestion hotspots. The solution has low resource footprint in terms of both battery usage on the 
sensing devices and the network bytes used for uploading data.  
When a user is not inside any hotspot zone, adaptive sampling conserves
battery power and reduces network usage, while ensuring that any new hotspots can be 
effectively identified. 
%A mix of sensors are used to automatically detect if a user is driving and to 
%trigger location identification and data upload to a central server. 
Our results show that our application consumes 40-80\% less energy than a periodic sampling system for different routes in our experiments, with similar accuracy of 
congestion information. The system can be used for a variety of applications such as automatic congestion 
alerts to users approaching hotspots, reliable end-to-end commute time estimates and effective alternate 
route suggestions.
\end{abstract}

%\begin{comment}
%
%- From a practical deployment challenges, and revisit design from addressing these as first priority, and make other metrics secondary.
%- why should a user upload data? esp. if battery drain and
%- why should we wait for a government to deploy the solution?
%- why should someone keep checking for incidents, which occasionally happen?
%- Most people are not that tech savvy for them to keep checking for directions, incidents, road traffic congestion, etc...
%-
%
%\end{comment}

% Add numbers from results.
% I have consciously left out mentioning using the tool for things other than traffic from our main claims

%% file: introduction.tex
\section{Introduction}
\label{sec:introduction}

%Mobile devices as sensors for a number of services. A key application area is traffic monitoring and management.

%Everyday, millions of man-hours and scarce fuel is being wasted as a result of poor transport 
%infrastructure and high density of vehicles during peak hours, especially in developing nations.
%The problem is exacerbated as several large organizations are collocated in IT parks or business
%centers leading to peak hour congestion. 

Traffic congestion is a problem that most urban locales struggle to grapple with.
In India, vehicle population has grown by over a 100 times in the last 50 years, while length of roads has 
increased by only 8 times~\cite{india-roads} during this period. It is estimated that 4.8 billion hours of time 
and 1.9 billion gallons of fuel were wasted due to congestion in 2011 in the US alone~\cite{us-traffic}. While 
governments strive to curtail congestion through various methods such as car-pooling incentives and congestion 
pricing, the onus has fallen on individual users to best cope with it.

%\begin{comment}
%-- past work has been focused on enabling technologies. We focus on addressing the practical challenges, 
%while building on part work.
%
%Lets not fix the congestion problem; lets fix the productivity problem. People who care will work around it. 
%People have different thresholds. %Allows auto-selection. Congestion problem automatically gets fixed. 
%Somewhat like Async.
%
%- low footprint solution: only areas that matter. Automatically figure out the hotspots.
%- privacy: automatically achieved at enterprise-level
%- distributed pub-sub system on enterprise and mobile side.
%\end{comment}

Recent research has focused on participatory sensing techniques for congestion detection as an 
alternative to infrastructure based systems that are often expensive and hard to maintain~\cite{centurymobile,thiagarajan2010cooperative,thiagarajan2009vtrack,mohan2008nericell}.
%Past work in this space has focused on several enabling technologies. A class of existing techniques attempt to 
%monitor congestion using sensors fitted on roads or on 
%cars~\cite{}. While such systems have been considerably successful in developed nations such as Japan,
%these tend to be expensive to set up and difficult to ensure sufficient participation. The other
%techniques to monitor congestion can be broadly classified as participatory sensing based
%approaches~\cite{}.
In participatory sensing, users willingly contribute information from sensors they already own (such as those in 
their cell phones), which can be aggregated and analyzed at a central server. Here, mobile 
phones essentially behave as mobile sensors uploading information from where they are at any point in time.
Unfortunately, proposed techniques using participatory sensing \textit{decouple collection 
of data from the use of the data}, and often use periodic sampling and uploading from mobile phones. 
Depending on the period of sampling, this often results in oversampling in dense regions and times 
than required, leading to increased battery and network bytes usage on user's mobile devices.

%Consequently, depending on the period of sampling, while the sampled data may be denser than required at some 
%places, it may not lead to sufficient samples to derive useful insights in other places. Where it is denser than 
%required, it leads to increased battery usage on mobile devices and consumes increased network bytes per user.

To this end, we formulate the following problem at an abstract level: given a set of mobile sensor devices, how do we collect GPS location information from them with low resource footprint, while 
deriving similar insights as uniform fine-grained sampling?
%These techniques typically require the user to manually start and stop sensing, and monitor the user's
%movement continuously and uniformly for this duration. Such methods are adequate when the user's movement is 
%smooth, without stark variations in speed, such as on a highway or in the absence of congestion. Virtual 
%triplines~\cite{} presents a privacy-preserving solution for ensuring that users can upload GPS data to 
%support traffic monitoring applications, while not disclosing their identity.
%In this paper, we argue that continuous uniform sampling approach used by past approaches is not really 
%required, and that
We build on the observation that {\em it is sufficient to selectively monitor certain key congestion hotspots}. Our approach is to \lq\lq{}couple\rq\rq{} data collection with the use or utility of the data collected, i.e., 
more data is collected in places and at times where there is more utility than at other places and times.
In fact, we observe through measurement studies that most of the speed variation with time on a person\rq{}s 
road trajectory is contributed by certain key hotspots.
Key challenges, however, are to define what a congestion hotspot is in a generic sense, and to identify a hotspot reliably.
Furthermore, congestion is a dynamic phenomenon, i.e., congestion hotspots may appear and disappear with time (e.g., due to accidents, construction, week-day office hours).

%These changes are tracked using an adaptive
%sampling approach introduced in this paper, that reduces the frequency of sampling a user's location as
%(s)he moves away from congestion hotspots.

We design and implement a system \name\ that consists of an Android application and a back-end analyzer that can automatically detect and monitor congestion hotspots. An adaptive sampling
approach ensures that the emergence of new hotspots can be detected quickly as soon as they 
\lq\lq{}become\rq\rq{} hotspots, while also conserving critical battery
power when the user is not driving within a hotspot zone. Further, a light-weight decision tree that uses a 
mix of sensors is used to detect if a user is driving, to automatically trigger location tracking using GPS 
and data upload to a central server. The central server utilizes the data uploaded by 
different smartphones and detects hotspots. 
%It manages user's location
%information in efficient data structures to allow spatio-temporal querying. This enables hotspot congestion 
%alerts to be selectively and automatically propagated to only those users approaching that particular
%hotspot, obviating the need  for active user participation.
We demonstrate that the system can make accurate estimates of user commute times and
suggest efficient alternate routes and commute start times, based only on information gathered at the
hotspots, consuming only half the battery power compared to uniform monitoring tools such as
Google Maps~\cite{googlemaps}. Our results are consistent across users of our Android application in a
large city in a developing nation, as well as in experiments using taxicab traces for San Francisco~\cite{comsnets09piorkowski}.  
%Our data analysis shows that reasonable estimates of end-to-end commute times 
%can be made based on monitoring just the congestion hotspots in real-time.

In summary, this paper makes two key contributions:
\begin{itemize}
\item Building on the observation that it is sufficient to monitor certain key congestion hotspots, we first explore how to define a hotspot in a generic sense, and then describe how to continuously and automatically  monitor and detect hotspots.
\item We design an adaptive monitoring technique for mobile phones that detects if the user is driving a vehicle, and monitors GPS location or speed more 
frequently as we approach closer to the {\em currently active} hotspots, and less frequently when away from the hotspots; the farther we are from the hotspot, the lower is the sampling frequency. This effectively couples the data collection to their use, ensuring energy efficiency.
\end{itemize}

The rest of the paper is organized as follows. Section~\ref{sec:motivation} formulates the problem, 
introduces the notion of hotspots, and discusses related work. Section~\ref{sec:design} describes the 
design and implementation of \name. Section~\ref{sec:analysis} provides analysis of our proposed approach. 
Section~\ref{sec:eval} presents performance evaluation of \name. Section~\ref{sec:conclusion} concludes the 
paper.

%%% I have downplayed the employee productivity angle
%%% Add more from results
%%% Emphasize what is new and different

%% file: motivation.tex
\begin{figure*}[tbh]
\begin{minipage}{2.5in}
\centerline{\includegraphics[height=1.4in]{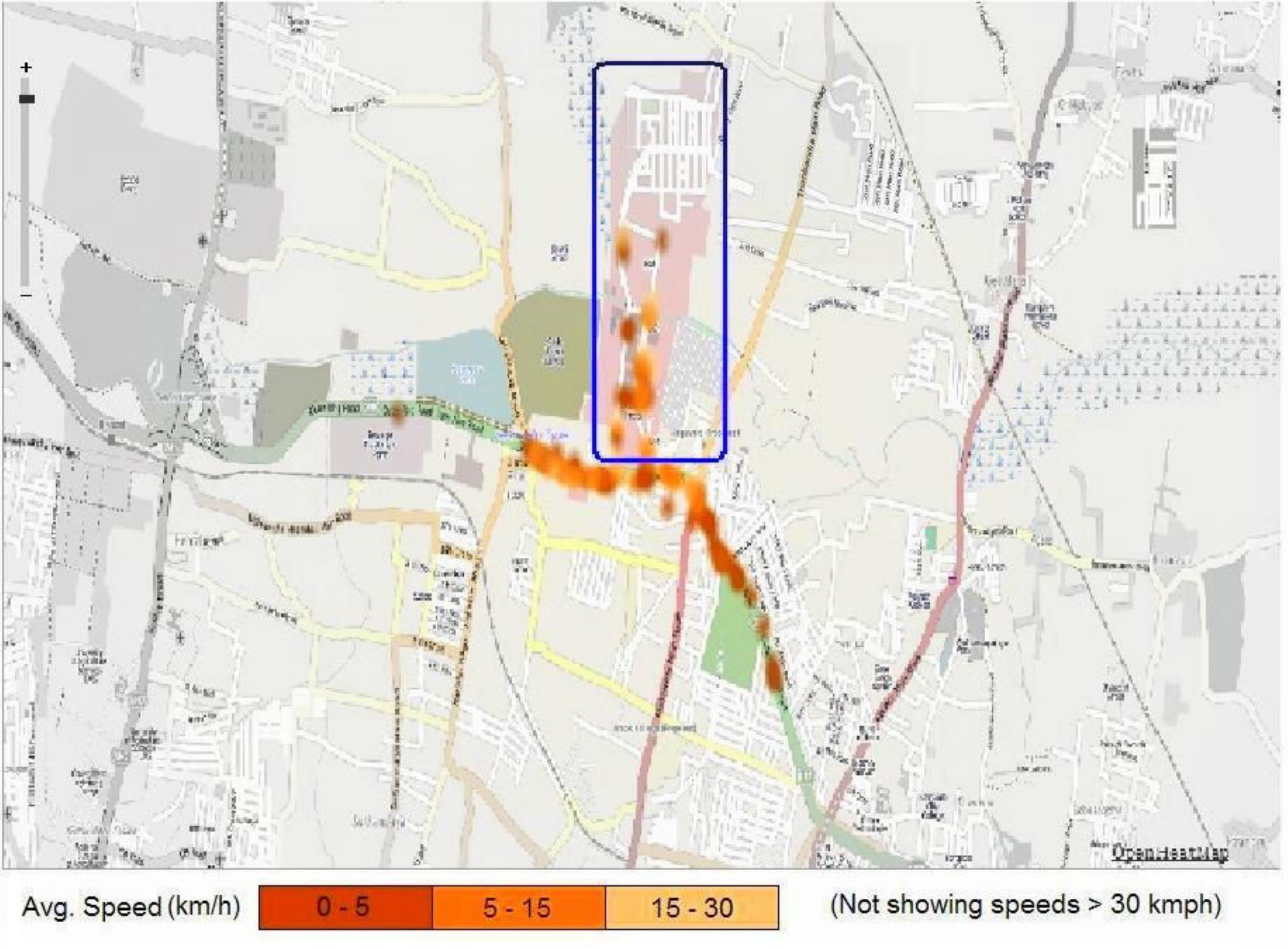}}
%\vspace{-0.2in}
\caption{\label{fig:Manyata-heat-map}Heat map of average speeds from commute routes of employees}
\end{minipage}
\quad
\begin{minipage}{1.6in}
\centerline{\includegraphics[height=1.4in]{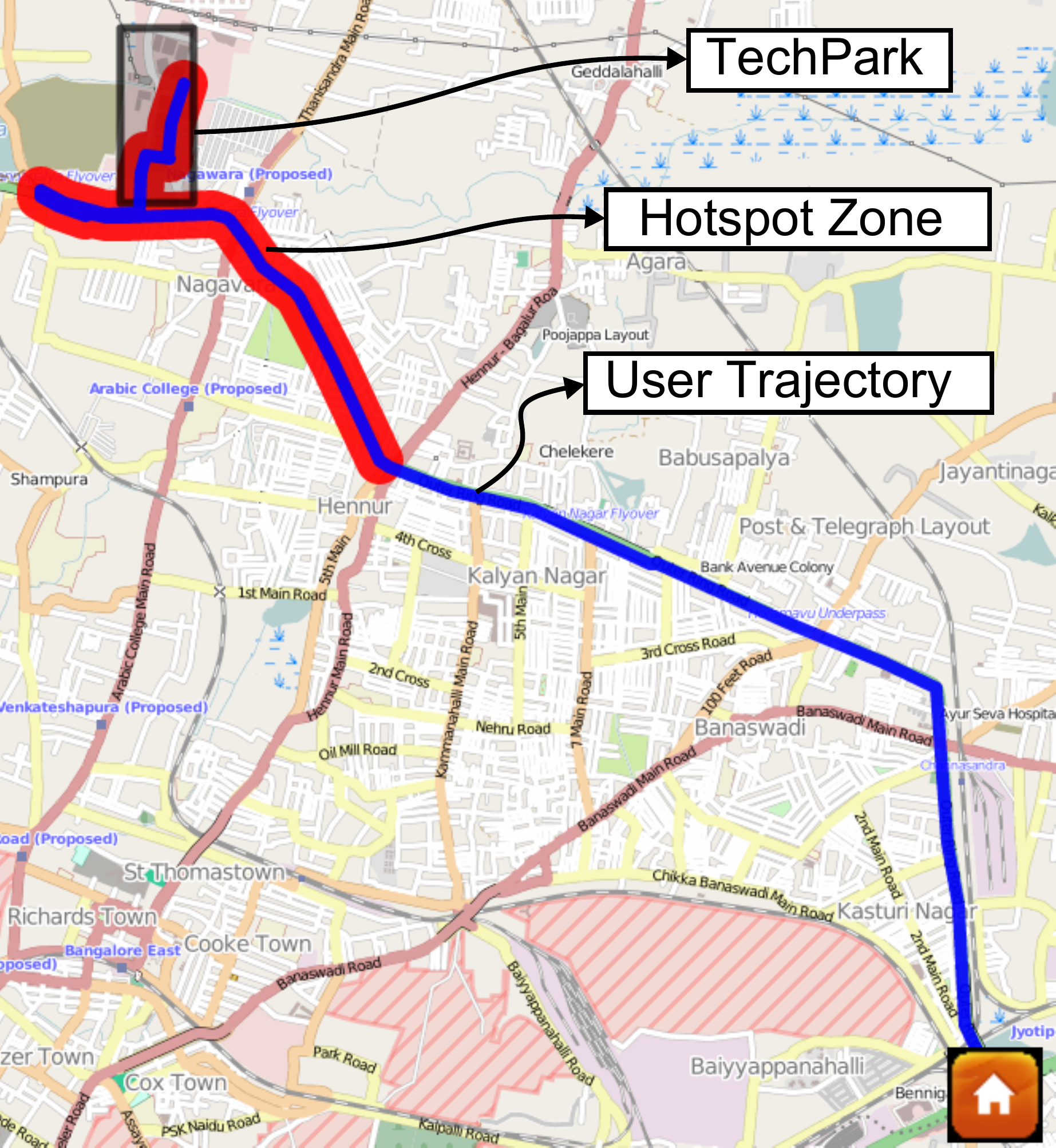}}
\caption{\label{fig:userTraj}User Trajectory and hot-spot regions}
\end{minipage}
\quad
\begin{minipage}{2.5in}
\centerline{\includegraphics[height=1.4in]{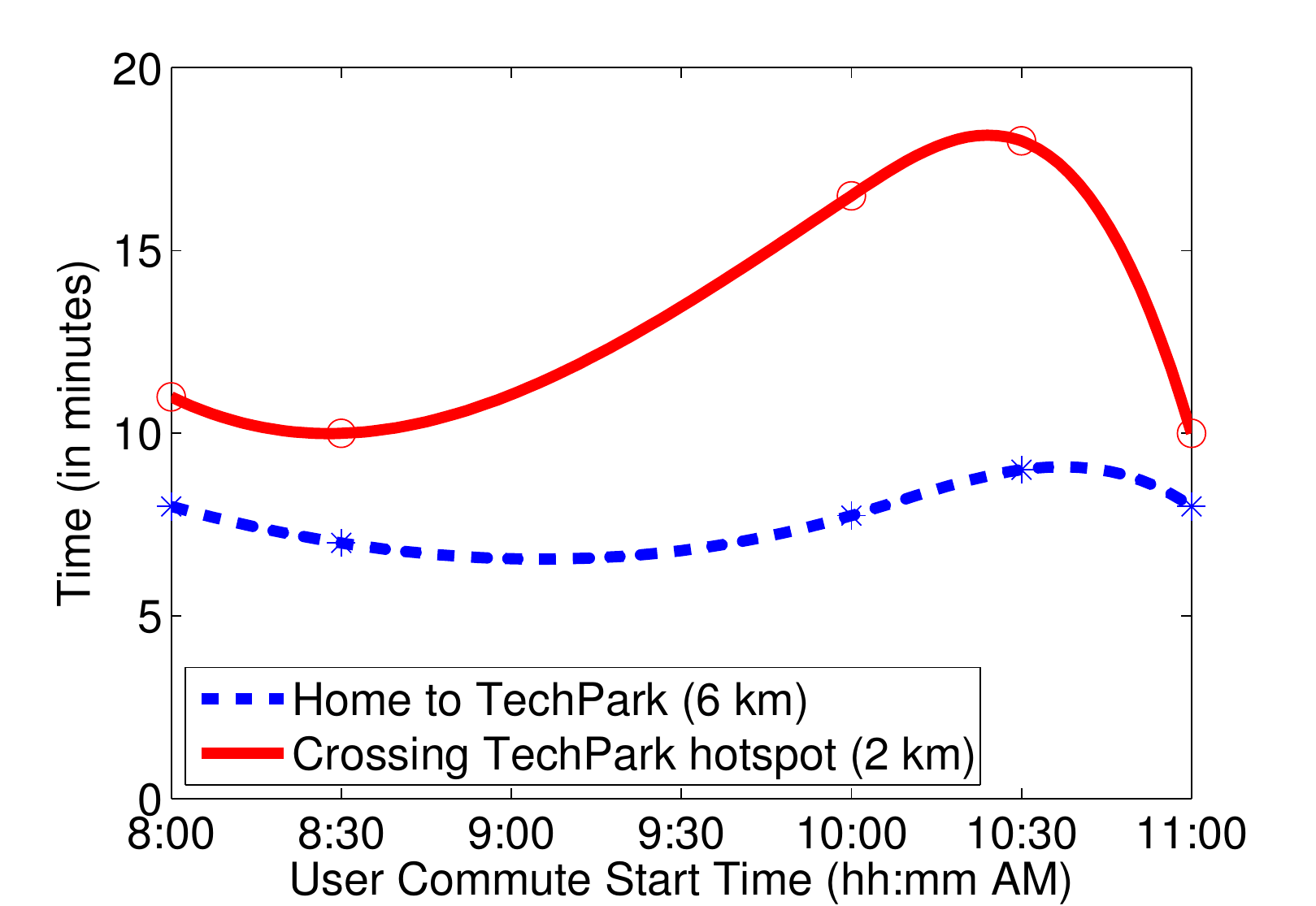}}
%\vspace{-0.2in}
\caption{\label{fig:split-time-Manyata-congestion}Split of time taken outside hotspot zone and within hotspot zone}
\end{minipage}
\vspace{-0.2in}
\end{figure*}

\section{Problem Formulation}
\label{sec:motivation}
In this section, we first motivate our solution approach through a simple experiment that shows the presence of traffic congestion hotspots. We then provide a formal definition of a hotspot, and discuss related work.

\vspace{-0.1in}
\subsection{Motivation}
Twelve employees from our organization used the smart phone application developed by us for two months, contributing location and speed information from their trips to and from office -- hereafter we refer to this office location as {\em TechPark} (marked with a rectangle in the Fig.~\ref{fig:Manyata-heat-map}).
The figure shows a heat map of average speed (points indicate only speeds less than \unit{30}{kmph}) on a road stretch just outside the TechPark across all these trips. Fig.~\ref{fig:Manyata-heat-map} shows that the average speed just outside the TechPark is mostly \unit{5-15}{kmph} during the commute hours, indicating the presence of a chokepoint or a hotspot. The speed on the same road segment is greater than \unit{30}{kmph} during afternoons and late nights.

%\begin{figure}[tbh]
%\vspace{-0.3in}
%\centering
%\includegraphics[height=2in,width=3in]{figs/TimeDivPraveenTrips.pdf}
%\vspace{-0.35in}
%\caption{Split of time taken outside hotspot zone and within hotspot zone.}
%\label{fig:split-time-Manyata-congestion}
%\vspace{-0.1in}
%\end{figure}

%\reminder{Plot x-axis to scale; Change Manyata to TechPark in figure}

%We now mark a 2 km road segment from the Manyata intersection (marked with arrow in Fgure~\ref{}) as a region of
%interest (hotspot zone), and divide the rest of the road segments that the employees traversed into similar
%regions.
%%To illustrate the nature and effect of congestion on commute times, we now consider the data from one particular 
To illustrate the nature and effect of congestion on commute times, we consider the data from a single
user (other users demonstrate similar trends). Over a period of two months, the user arrived at office each day at 
some time between \unit{8:00}{AM} and \unit{11:30}{AM} along the trajectory shown in Fig.~\ref{fig:userTraj}. 
We studied the commute time to traverse the \unit{2}{km} of his commute closest to the TechPark (referred to as 
{\em Hotspot Zone} in Fig.~\ref{fig:userTraj}), and compared this with the time taken to 
traverse the rest of the distance from his home, which was about \unit{6}{km}. Fig.~\ref{fig:split-time-Manyata-congestion} 
shows these two time-splits, namely, within the hotspot zone and outside the hotspot zone, for different times at which
the user entered the hotspot zone. Outside the hotspot zone, the user covered a distance of \unit{6}{km} in
about \unit{8}{minutes} on average. In contrast, the last \unit{2}{km} of his commute took more than 
\unit{15}{minutes} on average. Another interesting observation is that the variance in commute times is very low for 
the trip up to the hotspot, while the variance is considerably higher inside the hotspot zone. 

The above observations form the primary motivation for our work. We ask ourselves the following questions. How do we 
define and automatically detect such congestion hotspots? 
%As much of the variance in the overall commute times is due to the variance in traversing the hotspot zones, is it possible to monitor only the %hotspot zones in real-time, and yet achieve an accurate estimate of commute times? 
Can we develop a smart phone application that will require no active user participation, consume minimal battery power, 
and yet provide automatic and accurate congestion estimates? Is it possible to tune the system parameters to meet user's 
power budget constraints? We built a system that answers these questions in the affirmative.

We define a {\em congestion hotspot} as a geo-spatial region (specifically, a road segment of a certain length) with 
the following properties:
\begin{itemize}
\item $\operatorname{Current~average~speed} \leq \tau \times \operatorname{reference~speed}$, where $\tau$ is a 
predefined congestion threshold, and $\operatorname{reference~speed}$ is a reasonable achievable speed in the region. In 
our implementation, we use the maximum speed observed as the reference speed.
\item The region shows high temporal variance in average speed.
\item The congestion at all points within the hotspot region is similar at all times.
\end{itemize}

%Specifically, the geo-spatial region we use is a segment of a road of certain length. 
While several studies exist that define other notions of congestion zones~\cite{kerner2004three,kerner2004recognition}, 
we find that our simple definition is sufficient in practice as typical congestion situations are contained within this zone.

%% file: related.tex
\vspace{-0.1in}
\subsection{Related Work}
\label{sec:related}

GPS and other sensors are widely used to achieve low-cost traffic monitoring. In this section, we briefly discuss recent studies, and describe how we advance the state-of-the-art. 

\noindent\textbf{Participatory Sensing:}
Participatory traffic monitoring can be broadly classified into vehicle-mounted and smart-phone based sensing.  In vehicle-mounted sensing, location data is collected from moving objects that are mounted with GPS and other sensors~\cite{inrix,yoon2007surface}. Examples of such sensing include: (a) \textit{Green GPS} that computes fuel-efficient routes~\cite{ganti2010greengps}, (b) \textit{Pothole Patrol} accesses road surface conditions~\cite{eriksson2008pothole}, and (c) Taxi durations and fare estimation~\cite{balan2011real}.

In smart-phone based sensing, users are generally provided with an application that automatically collects data 
from the smart-phone sensors~\cite{burke2006participatory}. Mobile Millenium~\cite{centurymobile} and 
Nericell~\cite{mohan2008nericell} utilize multiple sensors on smart-phone, such as GPS, cellular connection and 
accelerometers, that detect traffic delays and congestion. A transit 
tracking system enables periodic GPS sampling if the user is moving in a vehicle, and then detects 
congestion is developed in~\cite{thiagarajan2010cooperative,thiagarajan2009vtrack}. 

%Similar to above studies, our work utilizes GPS, accelerometer, cellular and WiFi information on the smart-phone. However, we focus on accurate real-time congestion %detection and energy-aware sampling; our application intelligently decides when and what to sense such that the data transmitted is useful for accurate congestion %detection. 
Similar to the above studies, \name{} is a participatory sensing approach with minimal resource footprint, and 
minimal user intervention, which requires the user to primarily install and start the application once. Other 
applications providing similar services include Google Maps~\cite{googlemaps} and Waze~\cite{waze}, but both 
of them do not adapt based on the data utility. Moreover, they require significant user participation.
\rednote{Vinay: I think Sen's work and Signal Guru may be left out. Please add if you feel thats important}

{\bf Traditional Road Traffic Engineering Models:} There is a large body of work in traffic control systems for detecting congestion, such as three-phase theory~\cite{kerner2004three}, Forecasting of Traffic Objects (FOTO) and Automatic Tracking of Moving Traffic Jams (ASDA)~\cite{kerner2004recognition}. These systems typically require large amount of historical data to build reliable models, do not work for dynamic unpredictable hotspots, and require that traffic does not evolve significantly from the data for the models to remain valid.
%
%{\bf Traditional Road Traffic Engineering Models:} There is a large body of work in traffic control and intelligent traffic systems for detecting congestion, such as three-phase traffic theory (including free-flow, synchronized flow, and wide moving jam)~\cite{kerner2004three}, Forcasting of Traffic Objects (FOTO) and Automatic Tracking of Moving Traffic Jams (ASDA)~\cite{kerner2004recognition}. Model based systems typically require large amounts of historical data to build reliable models, do not work for dynamic unpredictable hotspots, and require that traffic does not evolve significantly from the data for the models to remain valid.

%% file: designTR.tex
\section{Design and Implementation}
\label{sec:design}

Given a road network, and a system of mobile devices traveling on the road network, the goal of \name\ is to (1) identify traffic congestion hotspots dynamically using data uploaded from the mobile devices, and (2) monitor the congestion hotspots continuously while being efficient in terms of battery and network resources utilized on the mobile devices. Several applications such as automatic congestion alerts to travelers, end-to-end commute time predictions, alternate route suggestions can be instantiated over \name.

For tractability, \name\ considers that a road network is broken down into road segments (including lines and curves)~\cite{osm}. Each road segment is a geospatial region that is tagged as a hotspot when it satisfies the conditions described in Section~\ref{sec:motivation}. To achieve the conflicting goals of maintaining reliability of tagging a region as a hotspot, and reducing the overhead of maintaining too many road segments, one needs to strike a tradeoff on the length of the road segments. For instance, longer segments (especially on highways) reduce the total number of road segments, but may be inaccurate in capturing congestion if only parts of the segments are congested during busy hours.
To balance this tradeoff, \name\ assumes that these segments are of 10s to 100s of meters. 
One side-effect of this tradeoff, however, is that adjacent road segments can be completely correlated at all times; in such scenarios, \name\ considers only one of the correlated segments as a representative hotspot, and avoids maintaining information or monitoring the other segments.

\subsection{Solution Overview}
\begin{figure}
%\vspace{-0.1in}
\centering
\includegraphics[scale=0.4]{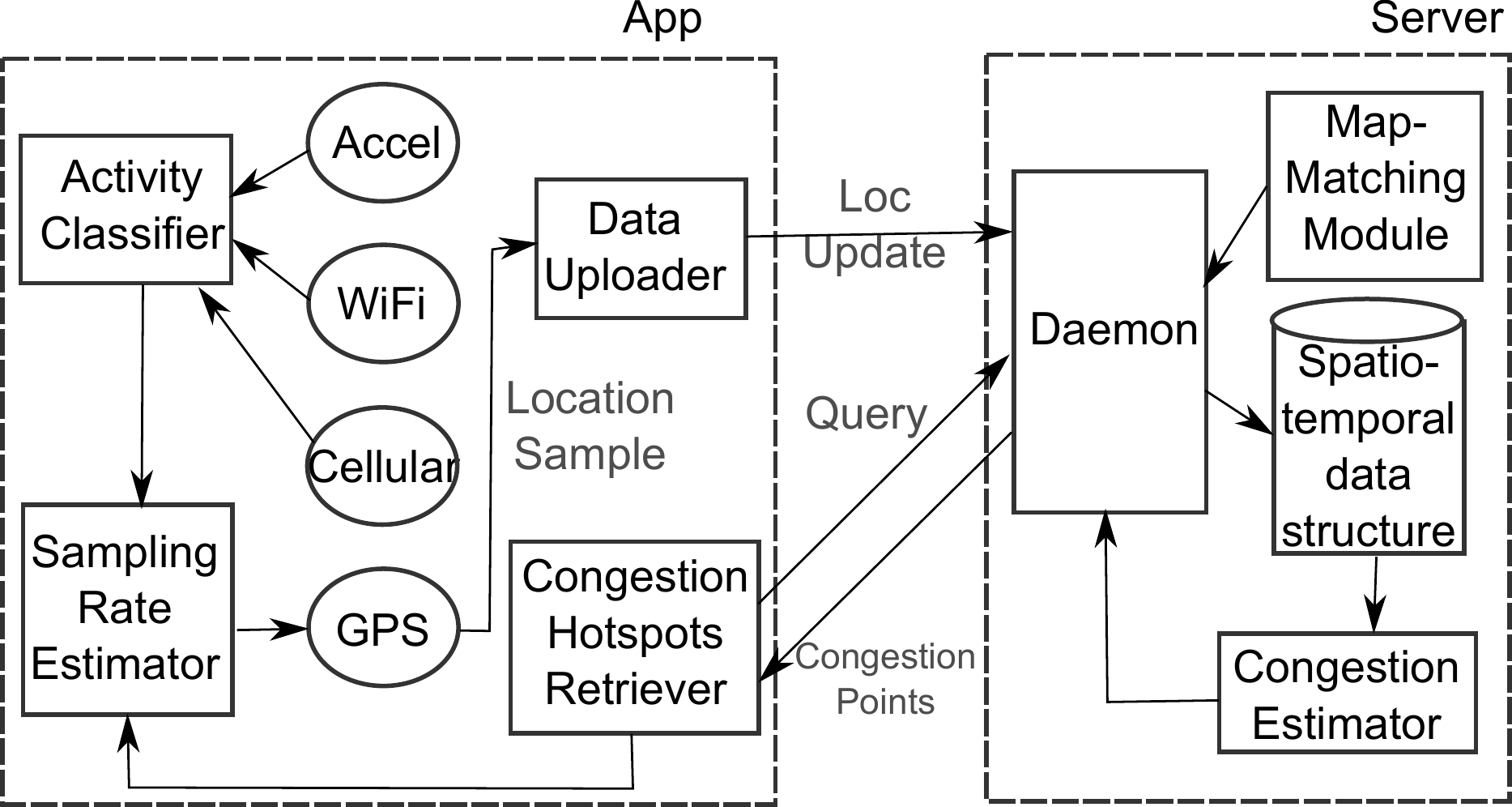}
%\vspace{-0.25in}
\caption{Overall system architecture.}
\label{fig:architecture}
%\vspace{-0.25in}
\end{figure}

We now describe the overall architecture and process of \name{}, pictorially shown in 
Figure~\ref{fig:architecture}. A smart-phone runs an application in the background to selectively sample GPS and 
upload the location data to the server in real time. On the phone, an activity classifier is invoked every 30 seconds to detect a user's activity. If the user is driving, it invokes Congestion Hotspot Retriever to obtain current congestion hotspot information from server, and Sampling Rate Estimator to adapt GPS sampling interval based on the nearest congestion hotspot. 
Once a location sample is obtained from GPS, it is uploaded to the server by the Data Uploader.
%Once a location sample is obtained from GPS, it is encoded in JSON format and uploaded to server by the Data Uploader over HTTP.
The server receives location updates and passes them to a Daemon. The Daemon validates each location and maps it to a point on the road as indicated by the Map-Matching Module. Map-matching corrects the GPS sampling errors and snaps the location samples to the road-paths traveled. In the current implementation, we use a Hidden Markov based map-matching~\cite{Newson2009}. User's information along with the updated location is placed into a spatio-temporal data structure, which is used by the Congestion Estimator to detect congestion hotspots.

\name\rq{}s functionality can be mainly categorized into two components: (1) detecting that a road segment is a hotspot, and (2) monitoring at the hotspot frequently, while also monitoring non-hotspots less frequently to be able to detect new hotspots.  We describe each components now.

\subsection{Hotspot Detection}
Hotspots can be detected using different approaches. A majority of them rely on parameters 
such as maximum speed limit on a road or number of lanes~\cite{kerner2004recognition}. 
While such parameters are easily available for roads in developed countries, they are either 
not defined, or changing over time, or not followed strictly in the developing countries. 
Further, these parameters are not sufficient to detect congestion caused by occasional incidents 
such as accidents or minor road works. Thus, to detect new congestion hotspots automatically, we 
do not assume knowledge of any such parameters. Instead, we use dynamic speed information 
received at the server for each road-segment over a long term to infer a {\em reference speed}, 
and view the current speed samples in relation to the reference speed.
%Congestion hotspots can be detected using different approaches. However, a majority of them rely on
%parameters such as maximum speed limit on a road, number of lanes, or maximum density of cars on a
%road-segment~\cite{kerner2004recognition}. While such parameters are easily available for roads in developed
%countries, they are either not defined, or changing over time, or not followed strictly in the developing
%countries. Thus, to detect congestion hotspots, we do not assume knowledge of these parameters. Instead,
%we use history of speed information received at the server for each road-segment to infer reference speed,
%and view the current speed samples in relation to the reference speed.

\name{} also recognizes that congestion is a dynamic phenomenon on a road 
segment, which varies based on the the spatial location of the road segment, 
time-of-the-day and unpredictable road incidents such as accidents and 
construction. We now describe how \name{} accounts for the spatio-temporal features 
of congestion and road dynamics to tag the road as a hotspot. 

\noindent\textbf{1. Accounting for temporal variability of speeds:}
The speeds on the roads vary based on temporal aspects such as time-of-the-day. 
To capture these temporal effects we divide a day into 48 equal \textit{time-bins}, 
each representing 30 minutes. For each road segment, we capture the number of 
samples received and average speeds. The server also maintains a maximum speed 
per segment across all times of the day. At each bin, we categorize congestion 
hotspots using low-, medium-, and high-congestion levels, similar 
to~\cite{kerner2004three}.

\noindent\textbf{2. Hotspot Marking:}
A time-bin of a road-segment can exist in three states: {\em uncongested}, {\em possible hotspot} and 
{\em hotspot}. We initially set all road-segment bins to {\em uncongested} state. As the user data trickles 
down to the server, the server updates the average speeds. Based on the aggregated values, the server 
classifies the road as a into one of the three states. 

Road segments may also experience flash-congestion due to incidents such as accidents or temporary 
obstructions. Automatic detection of flash hotspots is valuable since it can provide early warnings 
to users approaching a possible bottleneck or traffic regulators. To detect flash congestion, it is 
important to have a bounded sampling interval that is not larger than the time the user may take to 
cross the incident zone. When multiple recent samples with low speeds are obtained, a server tags the 
road-segment as {\em possible hotspot}. The server requests clients to increasingly sample road 
segments to conclude if the {\em possible hotspot} segment is a hotspot. We analyze the time required 
for flash-congestion detection in Section~\ref{sec:analysisCD}. 

\noindent\textbf{3. Separating hotspots from inherently low-speed roads:}
Different roads may have different speed signatures depending on the road size and road-surface conditions
such as speed-bumps and pot-holes. Thus, low average speeds do not necessarily imply that the road-segment
is a congestion hotspot.  We distinguish congestion hotspots from roads with inherently low speeds.

For each road-segment, we keep track of the maximum speed, $V_{\operatorname{max}}$ observed in the 
past. We then tag a segment as hotspot if the current average speed for the time-bin is lower than 
a threshold $T_{\operatorname{hot}} = 0.25 \frac{V_{\operatorname{max}}}{2}$. We set two more 
thresholds: $T_{\operatorname{med}}=0.5 \frac{V_{\operatorname{max}}}{2}$ and $T_{\operatorname{low}}=0.75 \frac{V_{\operatorname{max}}}{2}$. 
We tag the segment as medium- or low-congestion if the speeds are between
$(T_{\operatorname{hot}},T_{\operatorname{med}}]$ or $(T_{\operatorname{med}},T_{\operatorname{low}}]$,
respectively. We obtain these thresholds based on observations made from experiments at TechPark.

\subsection{Hotspot Monitoring}
For energy efficiency and minimizing the network usage, the application running on a smart phone detects when the user is mobile, and only then starts the monitoring activity, and it adapts the monitoring frequency based on the distance from the hotspot. If multiple hotspots are close to the mobile device, the closest hotspot determines the monitoring frequency. 

\subsubsection{Decision-Tree based Activity Detection}
Activity detection allows the system to be completely automated, thus not requiring active user 
participation; this is an important design goal for \name\ to ensure that people continue to use the application for a long time. Activity detection should accurately identify if a user is driving. False-positives in detection leads to excessive energy drain since the app will sample when the user is not driving (which is generally most of the day). False-negatives lead to loss of valuable data at the server side to detect hotspots. 

\begin{figure}
%\vspace{-0.1in}
\begin{center}
\mbox{
\subfigure[
~\label{fig:ActivityClassification}]{\includegraphics[scale=0.3]
{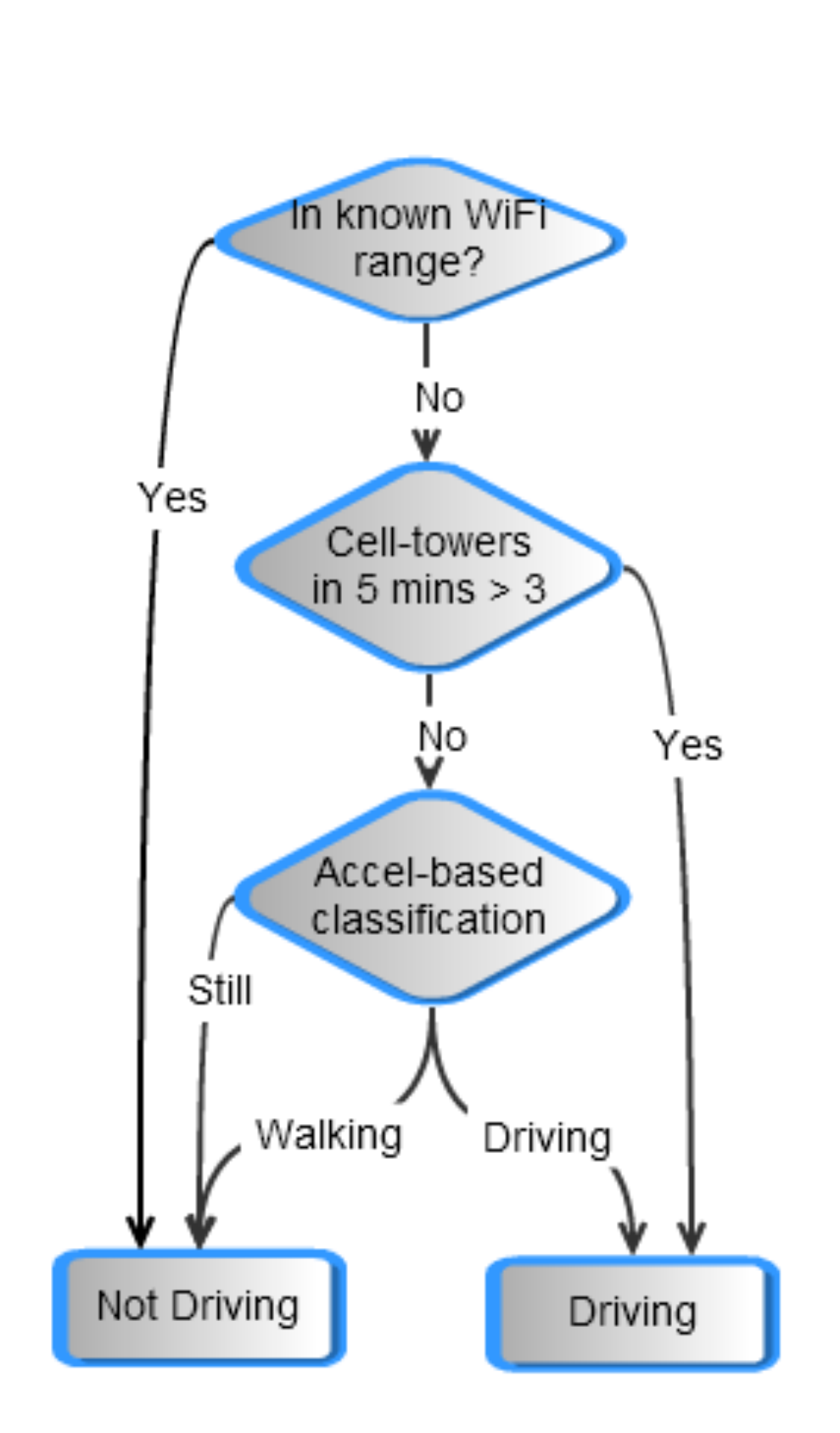}}
\quad 
\subfigure[
~\label{fig:AccelClassifier}]{\includegraphics[scale=0.3]
{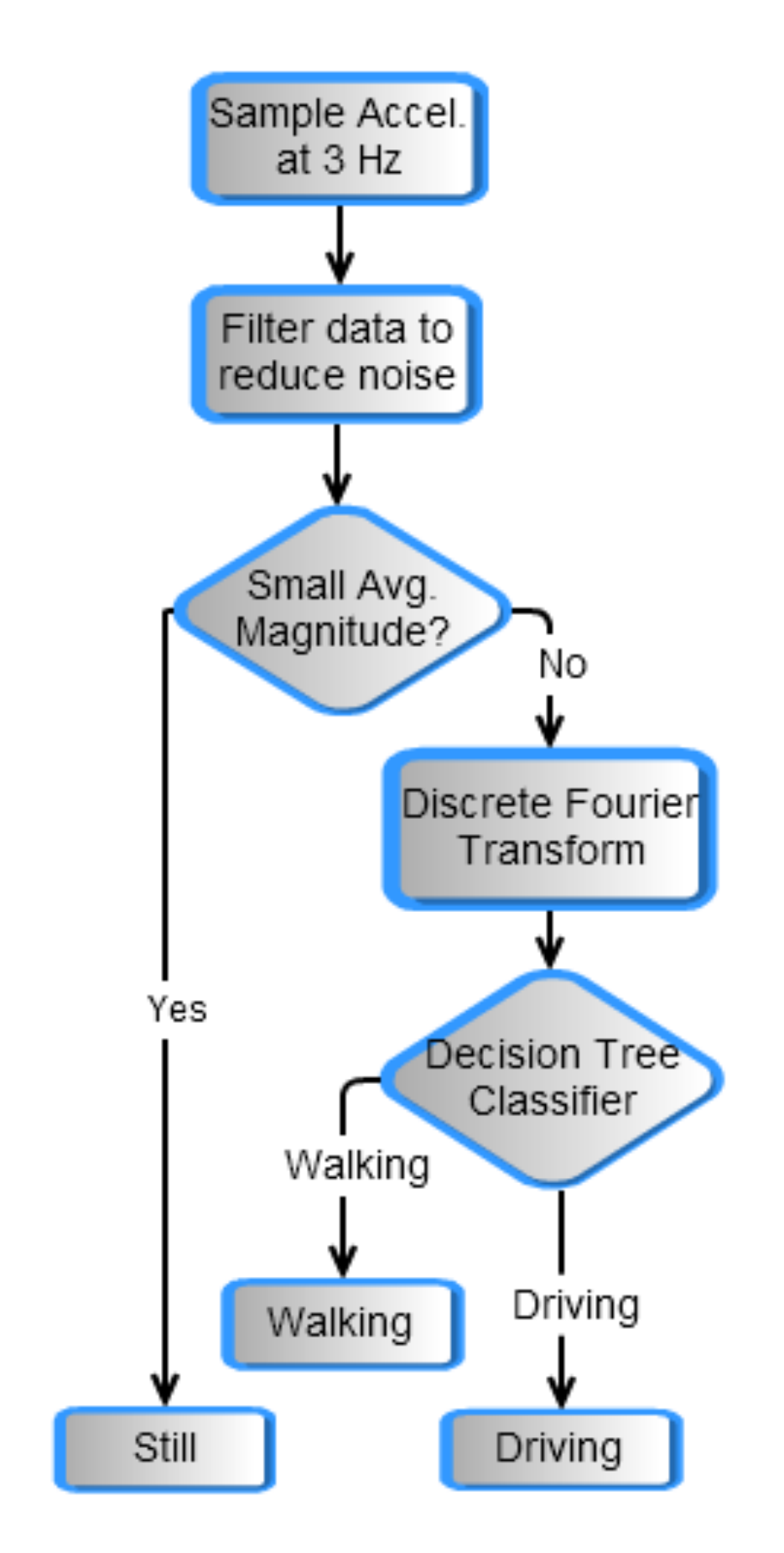}}}
\end{center}
%\vspace{-0.25in}
\caption{Activity classification: driving vs. not driving. \subref{fig:ActivityClassification} Activity classification using WiFi, GSM, and accelerometer sensors; \subref{fig:AccelClassifier} Accelerometer-based classification.}
%\vspace{-0.2in}
\end{figure}

We use an energy-aware combination of approaches for activity detection~\cite{thiagarajan2010cooperative,pier}, and switch-on and sample GPS only when the user is {\em most probably} driving. Figure~\ref{fig:ActivityClassification} shows our classification mechanism. We employ a decision tree 
where we first attempt to find clues of vehicular movement without sensing.
%
% for this, we use 
%WiFi and cell-tower information that is readily available on the phone. If the activity cannot be classified by such information, we use accelerometer-based classification, which is a low-energy sensing process.
%
%WiFi and cell-tower information is readily available 
%on smart-phones, and can be used to determine if the user is moving. Such intrinsic inference enables 
%low-energy classification since we avoid triggering any sensors. 
Specifically, we first check if the user is within the range of any known WiFi (office/home) access points. In such a case, the user is most likely not in a vehicle. Otherwise, we check the number of 
GSM cell-towers associated with in a given time window; if the number of towers is more than three in a short amount of time, we detect that the user is moving.
%; there is a high probability of crossing cell-towers if the user is moving in a vehicle. 
%If the number of cell-towers crossed in a given time is above a certain threshold, we classify that the user is 
%moving in a vehicle. 
%For example, based on our data samples at TechPark, we infer that the user is mobile if the user crosses 
%more than 3 cell-towers in 5 minutes.  
%
Otherwise, we use accelerometer readings to determine if the user is driving (as shown in 
Figure~\ref{fig:AccelClassifier})~\cite{thiagarajan2010cooperative}.

\subsubsection{Utility-aware Sampling}
\label{sec:adaptiveSampling}
Once the activity classifier detects that the user is moving in a vehicle, GPS sampling is enabled. However, the utility of GPS samples vary spatio-temporally, based on the current distribution of hotspots. Each phone periodically polls the server when the user is {\em driving}, to obtain a current list of congestion hotspots, where each congestion hotspot is defined in terms of its latitude and longitude based location, and approximate distance for congestion spread, $d_{\operatorname{in}}$. Based on this information, the phone selects the next GPS sampling interval $y$ as:

%\vspace{-0.15in}
\begin{eqnarray}
y \!\!\!\! & = \!\!\!\! & S_{\operatorname{min}}, \quad \quad \quad if\ d \leq d_{\operatorname{in}} \nonumber \\
\!\!\!\! & = \!\!\!\! & S_{\operatorname{min}} \! + \! \Big{(}1 \! - \! \frac{d_{\operatorname{in}}}{d}\Big{)} \! * \! (S_{\operatorname{max}} \! - \! S_{\operatorname{min}}), \ if \ d > d_{\operatorname{in}}
\label{eqn:adaptive-sampling}
\end{eqnarray}

\noindent where $d$ is the distance between the phone user and the closer of the end points of the {\em nearest} hotspot,
and $S_{\operatorname{min}}$ and $S_{\operatorname{max}}$ are the minimum and maximum sampling interval for GPS. If the user is within the congestion zone, then GPS is sampled at higher frequency. Otherwise, GPS is sampled adaptively, as the data becomes less useful. Values of $S_{\operatorname{min}}$ and $S_{\operatorname{max}}$ can be selected based on a power-budget for the application, which is described in more detail in Section~\ref{sec:analysis}. $S_{\operatorname{max}}$ provides a bound on the GPS sampling interval, which is useful in detecting new congestion hotspots.

The system supports querying about the congestion level at a given location and at a given time. 
The location data uploaded by the phone is recorded in a spatio-temporal data structure. We use
road network information to maintain spatial information~\cite{osm}. To
maintain temporal congestion information, we record the speed on roads at different times of the day. This 
allows the system to notify details such as hotspot location during a particular time of the day, and 
time required to cross it to the user.

We now describe the design goals of the system, certain key features, followed by an overview of the various system components.
%\vspace{-0.1in}
\subsection{Design Goals}

\begin{enumerate}
\item {\bf Require no active human participation:} Existing traffic monitoring tools, such as 
Google maps~\cite{googlemaps} or Waze~\cite{waze}, require active user participation to start or stop 
the application, query for directions, or to spatio-temporally annotate events. While such systems
are accurate, they are severely constrained as users are not always proactive. A fully
automated system enables large and ubiquitous user is likely to attract more users. However, such a 
automated system places various design challenges to accurately detect if the user is 
driving a vehicle and subsequently enable a location tracking service.

\item {\bf Energy efficiency without compromising accuracy:} Congestion detection requires 
GPS sampling and network transmission, which are energy intensive. \name{} should have minimal 
energy consumption so that the user is not required to frequently charge the smart-phone battery. 

\item {\bf Minimize network cost:} The sampled location data needs to be uploaded to a server 
in real-time, only if deemed pertinent. If the data exchange is high, users do not participate 
willingly in the data collection process -- especially when the mobile users pay per byte for 
cellular data connection.

\item {\bf Limited processing on the phones:} Phones are required to detect if the user is driving or not,
and accordingly start/stop GPS sampling and uploading. 
%Furthermore, they would need to predict utility of their sample and accordingly decide when to sample. 
However, phones have limited processing capacity. Thus, light-weight processing techniques is preferred 
with minimal loss of accuracy. Similarly, scalar-sensors are be preferred over audio/video sensors 
that require complex data processing operations on the phones.

\item {\bf Ability to answer spatio-temporal queries:} The system should be able to aggregate data to
detect and monitor congestion hotspots, and respond to spatio-temporal queries such as optimal time for
commute, providing congestion estimates for a particular location and/or time of day, etc.
\end{enumerate}

%\vspace{-0.1in}
\subsection{Design Features}
We next highlight key features of our design
%, namely (1) Utility-aware data sampling, (2) Energy-efficient adaptivity, (3) 
%Automatic congestion hotspot detection and monitoring, and (4) Spatio-temporal awareness.

\subsubsection{Utility-Aware Data Sampling}
Traditional sensor networks such as intrusion detection, and also existing traffic monitoring 
systems~\cite{googlemaps,waze} are ``sample, then analyze''-systems. Sensors uniformly collect 
the samples and later analyze the collected data. Such a system has the drawback of not analyzing
the possible {\em utility} of the data before sampling and transmitting. We adopt an efficient 
``analyze and sample simultaneously'' strategy where we analyze the potential utility of 
data before sampling, thus coupling the data collection with the possible utility of the data. In 
our system, the utility of data is ascertained using a few low-cost sensors
combined with knowledge of proximity of the user to current congestion hotspots.

%are deployed and used with the intention of uniformly
%{\it collecting samples and then analyzing} to see if any of the data was important. In this work, we adopt
%a different strategy, where we ``analyze and sample" simultaneously. We recognize that different data have
%different utility. Analyzing the potential utility of data before they are sampled tightly couples data
%collection with their use, and can make the sampling process efficient. In our system, the utility of data is ascertained using a few low-cost sensors
%combined with knowledge of proximity of the user to current congestion hotspots.

\subsubsection{Energy-Efficient Adaptivity}
Automatic congestion detection requires expensive GPS sampling. \name{} saves energy by using a 
combination of low-energy sensing using WiFi, GSM and accelerometer sensors. In addition,
we automatically assess the utility of each data sample based on how much the sample can
contribute towards accurately monitoring or detecting a hotspot. Our utility function considers 
various aspects such as distance from the nearest hotspot and last sampled time.

%The system should be able to automatically assess the utility of each data sample in order to be
%energy-efficient. Expensive GPS sampling is required only when the user is driving a vehicle.
%To detect if the user is driving, we employ an activity classification framework that uses a combination
%of low-cost sensors including WiFi, GSM cell tower information, and a 3-axis accelerometer. This low-cost
%activity classification mechanism allows the user to keep the application running permanently.
%
%Next, when the user is driving, we predict the utility of a new sample based on how much the sample can
%contribute towards accurately monitoring an existing hotspot or detecting a new hotspot. This utility
%estimation must be performed on the phone, and hence needs to be simple. We use two parameters to ascertain
%a new sample's utility: (1) distance of the user from the nearest congestion hotspot; (2) duration of time
%since the previous sample. The first parameter
%ensures that we sample more frequently when the user is closer to the congestion region, since our goal is
%to track the temporal-behaviour of a congestion hotspot in real-time. The second parameter, places a bound
%on the GPS sampling interval in order to detect new congestion hotspots.

\subsubsection{Automatic Congestion hotspot Detection and Monitoring}
Sampled location data from the smart-phone app is uploaded to the server in real-time. Not all of the GPS
data is accurate -- the location estimate could be erroneous or stale. Therefore, we first cleanse the GPS 
data~\cite{pier}, and accurate estimate the user location using techniques such as map-matching~\cite{Newson2009}.
%at the server side. \rednote{Are we doing any cleaning techniques at app-side? this would save energy.} 
%Various GPS data cleansing techniques have been proposed in literature~\cite{pier}.
%We use speed and timestamp information to remove erroneous GPS values.
%Next, we match the user's location on a road using an existing map-matching technique~\cite{}. The refined
%location is then stored in a spatio-temporal data structure, which is analysed to estimate congestion.
%Specifically, we term a location to be a congestion hotspot if the average speed observed at that location
%is significantly lower than the maximum speed observed over history.

% TODO: vikram: Bellow part is commented for now until we have time estimation graphs.
% Furthermore, we define congestion hotspots at the granularity of
% road-segments~\footnote{define roadsegment}, which allows us to accurately track % the spread of
% congestion on different roads. This is an important step in accurately estimating the time required to
% cross a congestion hotspot.

The system is also capable of detecting flash congestion, such as those created due to an accidents
or construction. Furthermore, the system distinguishes between congested roads and roads where 
vehicles cannot travel at high speeds due to, say, speed bumps and potholes.
%, due to the bounded GPS sampling rate. 
%system distinguishes lower-speeds due to congestion from low speeds due to small or bumpy roads.
%As one of the criteria for a congestion hotspot is to have temporal variations in speed, small
%or bumpy road segments that have a uniformly low speed do not fit this criteria and are not categorized as 
%congestion hotspots.

\subsubsection{Spatio-Temporal Awareness}
The system supports querying about the congestion level at a given location and at a given time. 
The location data uploaded by the phone is recorded in a spatio-temporal data structure. We use
road network information to maintain spatial information~\cite{osm}. To
maintain temporal congestion information, we record the speed on roads at different times of the day. This 
allows the system to notify details such as hotspot location during a particular time of the day, and 
time required to cross it to the user.

%For spatial representation, we construct a road-network graph, where each vertex represents smaller portions of roads (called as road-segments). Connectivity among vertices maps to the connectivity on roads. For each road segment, we maintain a list of observed vehicle speeds in time-order. This allows the system to notify details such as hotspot location and time required to cross it to the user.

%\reminder{Briefly describe the spatio-temporal data structure that makes this possible. Details can be
%covered in the implementation}

%The overall system involves a smart-phone based data collecting application, which selectively collects and
%uploads GPS locations to the server. Server stores the location traces into a spatio-temporal data-structure,
%and detects and monitor congestion hotspots.
%The key aspect in which the system differs from the existing systems is its utility-based sampling. The
%existing systems~\cite{googlemaps,waze} monitor vehicle-locations continuously and upload the data to infer
%traffic condition. Such systems may result in either over-sampling or under-sampling, since they do not
%account for the utility of data. Instead, we need an approach of sampling and uploading data only when
%the utility of the data is high. Data utility must be determined locally on the phone, which can be used
%to decide appropriate GPS sampling rate. Towards these goals, our system supports the following operations:
%(1) Energy-Efficient Sampling (2) Congestion hotspot detection and monitoring; (3) Spatio-Temporal awareness.

%% file: analysisTR.tex
\section{Analysis}
\label{sec:analysis}
In this section, we analyze the utility-aware adaptive sampling and new hotspot detection approaches. 

\subsection{Energy budget for Adaptive Sampling}
\label{sec:analysisAS}
%We have modeled 
We model the energy consumption of adaptive sampling, and then use the analysis to choose 
appropriate values for the parameters in our adaptive sampling approach, that will
ensure that the application does not consume more energy than stipulated. 

%We describe the intuition and results of the model here. Detailed derivation is present in the technical report~\cite{rtChokeTechRep}.

\begin{table}[ht]
%\vspace{-0.1in}
\begin{small}
\begin{center}
\begin{tabular}{|c|c|} \hline
Symbol & Description \\ \hline
$d_{\operatorname{in}}$ & Distance inside congestion zone \\
$d_{\operatorname{out}}$ & Distance outside congestion zone \\
$v_{\operatorname{in}}$ & Avg. speed inside congestion zone \\
$v_{\operatorname{out}}$ & Avg. speed outside congestion zone \\
$S_{\operatorname{min}}$ & Min. sampling interval \\
$S_{\operatorname{max}}$ & Max. sampling interval \\
$e_{\operatorname{in}}$ & Energy per sample inside congestion zone \\
$e_{\operatorname{out}}$ & Energy per sample outside congestion zone \\ \hline
\end{tabular}
\end{center}
\end{small}
%\vspace{-0.15in}
\caption{Notation used in analysis.}
\label{tab:notation}
%\vspace{-0.1in}
\end{table}

The notation used in the analysis is presented in Table~\ref{tab:notation}. We make the following 
assumptions for our analysis. While these make the analysis simplistic, it helps us make 
intelligent choices for the values of parameters in our adaptive sampling approach. We consider a user who starts from 
the congestion hotspot and travels along
a straight path to their destination covering a distance $d_{\operatorname{in}}$ within the congestion zone first and then a distance $d_{\operatorname{out}}$ outside the congestion zone (the analysis is equally applicable for users traveling towards the congestion hotspot). We assume that the average speed within the congestion zone is $v_{\operatorname{in}}$
and that the user moves with a uniform speed of $v_{\operatorname{out}}$ outside the congestion zone. Within the congestion
zone, the application samples GPS values with a constant sampling interval of $S_{\operatorname{min}}$. Outside the 
congestion zone, we use the sampling interval as shown in Equation~\ref{eqn:adaptive-sampling}.
Let $x = d - d_{\operatorname{in}}$ denote the distance covered by the user outside the congestion zone at the instant
a sample was taken. Simplifying Equation~\ref{eqn:adaptive-sampling}, we get the sampling interval $y$:

%\vspace{-0.15in}
\begin{equation}
y = \frac{S_{\operatorname{max}} x + d_{\operatorname{in}}S_{\operatorname{min}}}{d_{\operatorname{in}} + x}
\label{eqn:adaptive-sampling-alt}
\end{equation}

%\vspace{-0.1in}
%\begin{equation}
%y = S_{\operatorname{min}} + \Big{(} \frac{x}{d_{\operatorname{in}} + x} \Big{)} (S_{\operatorname{max}} - S_{\operatorname{min}})
%\label{eqn:adaptive-sampling}
%\end{equation}
%\noindent where $x$ is the distance covered outside the congestion zone at the instant a sample was taken,
%and $y$ is the sampling interval until next sample. 

We now bound the total energy used by the application during the course of the user's trip. The number of 
samples taken within the congestion zone is,

%\vspace{-0.15in}
\begin{equation*}
n_{\operatorname{in}} = \frac{d_{\operatorname{in}}}{v_{\operatorname{in}}} \frac{1}{S_{\operatorname{min}}}
\end{equation*}

We next determine an upper bound on the number of samples outside the congestion zone. Let $x_i$ denote
the distance covered up to the $i^{th}$ sample outside the congestion zone and $y_i$ denote the sampling 
interval after the $i^{th}$ sample. It is safe to assume that the first sample outside the 
congestion zone is taken right at the 
edge of the zone, so $x_1 = 0$. Therefore, $y_1 = S_{\operatorname{min}}$. We compute the distance covered and sampling 
interval for the subsequent samples as follows.

%\vspace{-0.15in}
\begin{eqnarray*}
x_2 = v_{\operatorname{out}} S_{\operatorname{min}}; \quad y_2 & = & \frac{S_{\operatorname{max}} x_2 + d_{\operatorname{in}} S_{\operatorname{min}}}{d_{\operatorname{in}} + x_2} \\
& = & S_{\operatorname{min}} \Big{(} \frac{d_{\operatorname{in}} + v_{\operatorname{out}} S_{\operatorname{max}}}{d_{\operatorname{in}} + v_{\operatorname{out}} S_{\operatorname{min}}} \Big{)} 
\end{eqnarray*}

%\vspace{-0.15in}
\begin{eqnarray*}
x_3 = x_2 + v_{\operatorname{out}} y_2 = v_{\operatorname{out}} S_{\operatorname{min}} \Big{(} 1 + \frac{d_{\operatorname{in}} + v_{\operatorname{out}} S_{\operatorname{max}}}{d_{\operatorname{in}} + v_{\operatorname{out}} S_{\operatorname{min}}} \Big{)}
\end{eqnarray*}

Substituting for $x_3$ in Equation~\ref{eqn:adaptive-sampling-alt} and simplifying,

%\vspace{-0.15in}
\begin{eqnarray*}
\!\!\!\!\! y_3 \!\!\!\! & = \!\!\!\! & S_{\operatorname{min}} \frac{ (d_{\operatorname{in}} + v_{\operatorname{out}}S_{\operatorname{max}})^2 + d_{\operatorname{in}}v_{\operatorname{out}}S_{\operatorname{min}} + v_{\operatorname{out}}^2 S_{\operatorname{min}}S_{\operatorname{max}}}{ (d_{\operatorname{in}} + v_{\operatorname{out}}S_{\operatorname{min}})^2 + d_{\operatorname{in}}v_{\operatorname{out}}S_{\operatorname{min}} + v_{\operatorname{out}}^2 S_{\operatorname{min}}S_{\operatorname{max}}} \\
\!\!\!\!\! & > \!\!\!\! & S_{\operatorname{min}} \frac{ (d_{\operatorname{in}} + v_{\operatorname{out}}S_{\operatorname{max}})^2 }{ (d_{\operatorname{in}} + v_{\operatorname{out}}S_{\operatorname{min}})^2 }
\end{eqnarray*}

\noindent as $\frac{a_1 + b}{a_2 + b} > \frac{a_1}{a_2}$, for $a_1, a_2, b > 0$. Using these as basis 
step, it can be shown using induction on the number of samples $n_{\operatorname{out}}$ that,

%\vspace{-0.15in}
\begin{eqnarray}
\!\!\!\!\! x_{n_{\operatorname{out}}} \!\!\!\! & \geq \!\!\!\! & v_{\operatorname{out}} S_{\operatorname{min}} \Big{[} 1 \! + \! \Big{(} \frac{d_{\operatorname{in}} + v_{\operatorname{out}} S_{\operatorname{max}}}{d_{\operatorname{in}} + v_{\operatorname{out}} S_{\operatorname{min}}} \Big{)} \! + \! \Big{(} \frac{d_{\operatorname{in}} + v_{\operatorname{out}} S_{\operatorname{max}}}{d_{\operatorname{in}} + v_{\operatorname{out}} S_{\operatorname{min}}} \Big{)}^2 \nonumber \\
\!\!\!\! & \!\!\!\! & + \ldots + \Big{(} \frac{d_{\operatorname{in}} + v_{\operatorname{out}} S_{\operatorname{max}}}{d_{\operatorname{in}} + v_{\operatorname{out}} S_{\operatorname{min}}} \Big{)}^{n_{\operatorname{out}}-2} \Big{]} \nonumber \\
\!\!\!\!\! & = \!\!\!\! & v_{\operatorname{out}} S_{\operatorname{min}} \frac{ \Big{(} \frac{d_{\operatorname{in}} + v_{\operatorname{out}} S_{\operatorname{max}}}{d_{\operatorname{in}} + v_{\operatorname{out}} S_{\operatorname{min}}} \Big{)}^{n_{\operatorname{out}} - 1} - 1}{\frac{d_{\operatorname{in}} + v_{\operatorname{out}} S_{\operatorname{max}}}{d_{\operatorname{in}} + v_{\operatorname{out}} S_{\operatorname{min}}} - 1}
\end{eqnarray}

In order to bound the number of samples needed to cover a distance $d_{\operatorname{out}}$ outside the congestion zone,
we require that $x_{n_{\operatorname{out}}-1} < d_{\operatorname{out}} \leq x_{n_{\operatorname{out}}}$.
Substituting $r = \frac{d_{\operatorname{in}} + v_{\operatorname{out}} S_{\operatorname{max}}}{d_{\operatorname{in}} + v_{\operatorname{out}} S_{\operatorname{min}}}$ and solving for $n_{\operatorname{out}}$,

%\vspace{-0.15in}
\begin{equation}
n_{\operatorname{out}} = 1 + \Big{\lceil} \log_{r} \Big{(} 1 + \frac{d_{\operatorname{out}} (r - 1)}{v_{\operatorname{out}} S_{\operatorname{min}}} \Big{)} \Big{\rceil}
\end{equation}

Although, there are other computations that execute periodically, such as evaluating the decision tree 
to determine if the user is driving, for simplicity we assume that $e_{\operatorname{in}}$ and $e_{\operatorname{out}}$, the average 
energy values consumed per sample inside and outside the congestion region, respectively, is uniform 
across all samples (this includes energy used for sensing, computation, and communication). Hence, the 
total energy used by the application for the entire trip can be determined as,

%\vspace{-0.15in}
\begin{equation}
E = n_{\operatorname{in}} e_{\operatorname{in}} + n_{\operatorname{out}} e_{\operatorname{out}}
\label{eqn:energy-function}
\end{equation}

\begin{figure}[htbp]
%\vspace{-0.15in}
\begin{center}
\begin{turn}{270}
\includegraphics[width=4.5cm]{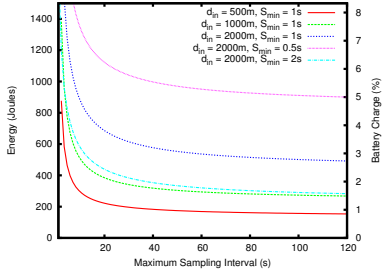}
\end{turn}
%\vspace{-0.1in}
\caption{Total energy of app from analysis.}
% for different values of congestion zone radius and minimum, maximum sampling intervals
\label{fig:energy-analysis}
\end{center}
%\vspace{-0.3in}
\end{figure}

Based on average values from our local city experiments, we obtained the average speed 
within the congestion zone as $v_{\operatorname{in}} = 2 m/s$ ($16.67 kmph$), and the average speed outside the congestion 
zone as $v_{\operatorname{out}} = 12.5 m/s$ ($45 kmph$). The energy consumed per sample were obtained as, 
$e_{\operatorname{in}} = 0.4J$ and $e_{\operatorname{out}} = 7J$ (note that $e_{\operatorname{out}}$ is for a much larger time period as samples 
are obtained less frequently). We plot the energy as a function of the maximum sampling interval 
(as derived in Equation~\ref{eqn:energy-function}), for different values of the congestion zone radius
$d_{\operatorname{in}}$ and different minimum sampling intervals in Figure~\ref{fig:energy-analysis}, assuming that the 
total distance covered by the user in the trip is $d_{\operatorname{in}} + d_{\operatorname{out}} = 10km$.

We note that as the maximum sampling interval is increased, the energy needed by the application reduces rapidly, stabilizing beyond values of 60s. Further, we note that the energy consumption is sensitive to the  congestion zone radius, as this is where a majority of the samples are obtained. For the same reason, the sampling interval within the congestion zone also has a considerable affect on the energy consumed. 
For our experiments, we used $d_{\operatorname{in}} = 2000m$ as we observed that most congestion situations were contained 
within this stretch. The smart phones used in our experiments had battery energy of about 18000J and we set a goal for \name\ that the application should not use more than 3\% of the battery charge, which amounts to 560J. Based on this constraint, we choose $S_{\operatorname{min}} = 1s$ and $S_{\operatorname{max}} = 120s$ for our experiments. 

\subsection{Hotspot Detection}
\label{sec:analysisCD}
Consider a road segment $R$ where a congestion builds up due to an incident, such as an 
accident, at time $t=0$. We compute the time taken for the server to detect that the segment might be 
a possible hotspot. 
%Let  $\rateEntry$ be the arrival rate of the cars into the segment $R$, and 
Let $\currSamplingInterval$ be the current sampling interval on $R$ given the location of existing 
hotspots (derived from Equation~\ref{eqn:adaptive-sampling}). 
%Note that $\currSamplingInterval=S_{\mymax}$ if the given $R$ is far from any existing hotspot.

We assume that the server tags the segment as a possible hotspot if it receives $\numSamplesReqd$ 
samples. 
%Generally, we set $\numSamplesReqd=30$ by applying central limit theorem. This is a valid 
%assumption since most of the cars travel at similar speeds in a possible congestion hotspot, 
%and thus the distribution can be assumed to be normally distributed. 
We also assume that the communication of speeds from the application to the server is error-free. 
We first compute the number of vehicles that are on the road segment as a function of time, and 
then estimate the time required for the server to receive $\numSamplesReqd$ samples.

\noindent\textbf{1. Vehicle build-up function:} We denote the rate at which the vehicles enter 
the road segment $R$ by $\rateEntry$ vehicles per unit time. Let $\rateExit$ be the rate 
at which the traffic exists the hotspot. We assume that $\rateExit \le \rateEntry$ at $R$, and hence 
the queue dynamics leads to infinite buildup until the traffic regulators defuse the congestion by 
external means. Our aim is to estimate the time required to detect congestion buildup, and thus provide 
a framework to notify flash congestion to travelers and regulators. 

The number of vehicles that accumulate on $R$ in time $t$ is given by $\numCars(t) = (\rateEntry - \rateExit) t$;
the congestion buildup increases linearly.
\\\noindent\textbf{2. Number of samples at time $t$:} We now estimate the number of samples that
the server receives at a given time $t$. Since one car sends the sample every $\currSamplingInterval$ 
time units, the expected number of samples at time $t$ from $\numCars(t)$ is 
$\frac{\numCars(t)}{\currSamplingInterval}$. Hence, over a period $[0,t]$, the total number 
of samples received at the server is defined by
%\vspace{-0.1in}
\begin{eqnarray}
\numSamplesAtTime(t) = \int_0^t \frac{\numCars(x)}{S} dx = \frac{(\rateEntry - \rateExit) t^2}{2S}.
\label{eqn:numSamplesAtTime}
\end{eqnarray}

The time required for the server to collect $\numSamplesReqd$ samples to detect if $R$ is a possible 
hotspot can be determined by equating $\numSamplesAtTime(t)=N$. Hence, 
%\vspace{-0.1in}
\begin{eqnarray}
\timeReq&=& \sqrt{\frac{2 \numSamplesReqd \currSamplingInterval}{\rateEntry - \rateExit}}~~.
\end{eqnarray}

\noindent Hence, the time required to detect a road segment as a hotspot at the server decreases drastically, 
as a square root function of differential of the rates at the segment. 

\begin{figure}
%\vspace{-0.1in}
\centering
\includegraphics[scale=0.4]{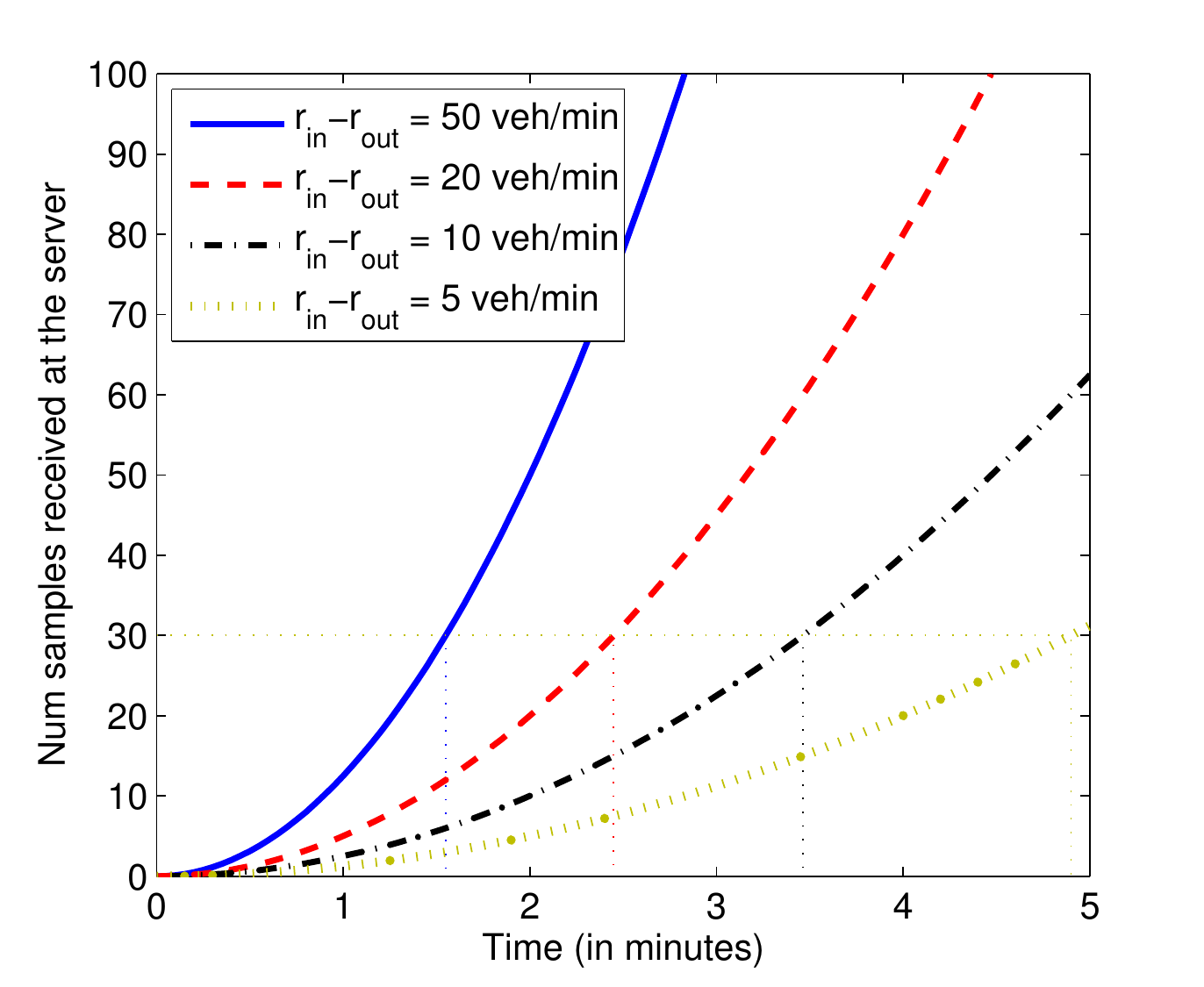}
%\vspace{-0.15in}
\caption{Time required to detect a hotspot decreases rapidly as $\rateEntry - \rateExit$ increases}
\label{fig:flashDetect}
%\vspace{-0.1in}
\end{figure}

Figure~\ref{fig:flashDetect} shows the time required to detect a hotspot where $\numSamplesReqd=30$, 
$\currSamplingInterval=S_{\mymax}=2$ minutes, and for different $\rateEntry$ and $\rateExit$. Rates are 
measured in vehicles per minute, and are derived from existing traffic literature~\cite{jain2012road}. 
As the difference in the entry and exit rate increases, the detection time for congestion 
decreases rapidly. Once the server detects the congestion hotspot, regular adaptive sampling can 
be resumed.

%% file: experiments_blr.tex
\section{Experimental Evaluation}
\label{sec:eval}
We evaluate \name{} using two sets of data. The first set was obtained from a pilot study of 12 smart-phone
users commuting to and from their work place at TechPark, located in a major city in a developing country,
over a period of almost 2 months. The second set was traces of taxicab location information in San Francisco
city~\cite{comsnets09piorkowski}.

\vspace{-0.1in}
\subsection{Congestion at TechPark}
In Section~\ref{sec:motivation}, Figures~\ref{fig:Manyata-heat-map} and~\ref{fig:split-time-Manyata-congestion},
showed that the roads close to TechPark experienced much lower average speeds during peak hours than roads
that were farther away and that the observed speeds varied significantly with time of day. In our evaluation,
we show: (1) \name{} can accurately detect congestion hotspots that exhibit low average speeds as well as
temporal variation, (2) Effectiveness of adaptive-sampling compared to continuous sampling, and (3)
Comparison of the energy footprint of our application with that of Google Maps with traffic updates enabled.
For these experiments, we set $d_{in} = 2000m$, $S_{\operatorname{min}} = 1s$, and $S_{\operatorname{max}} = 2min$, adhering to the analysis
presented in Section~\ref{sec:analysis}.
%For large-scale experiments, we evaluate our system using the San Francisco cabs dataset.

\vspace{-0.05in}
\subsubsection{Real-time Congestion Detection}

\begin{figure}
\vspace{-0.1in}
\centering
\includegraphics[scale=.22]{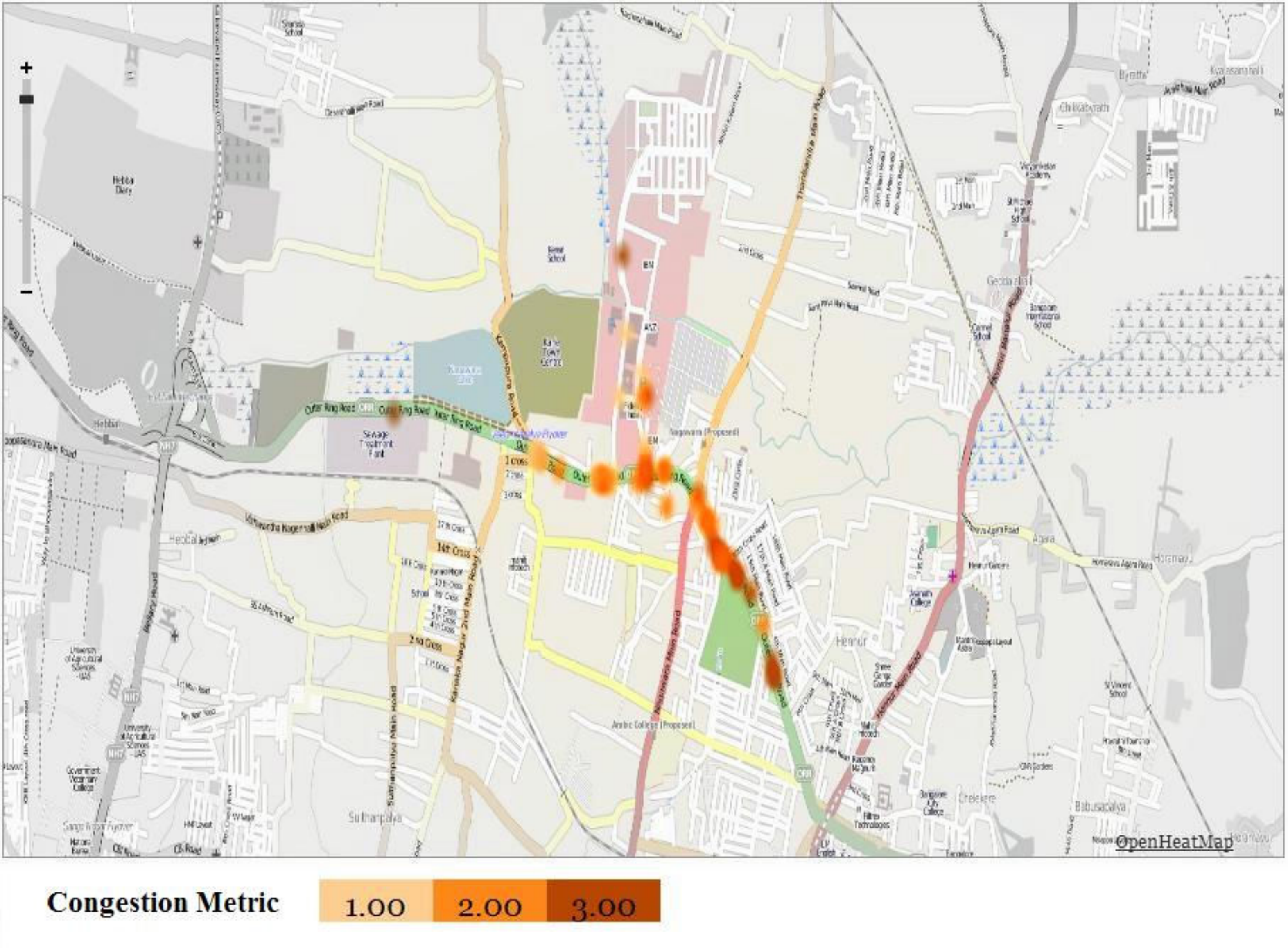} %BLR_Manyata_12Cars_9AM_CM.pdf}
\vspace{-0.1in}
\caption{Congestion detection at TechPark.}
\label{fig:BLR_Manyata_10_15AM_CM}
\vspace{-0.1in}
\end{figure}

The TechPark region (highlighted with a blue rectangle in Figure~\ref{fig:Manyata-heat-map} in
Section~\ref{sec:motivation}) hosts multiple
organizations employing more than $100,000$ people. As noticeable in the figure, the average speeds just
outside TechPark are quite low during peak hours due to severe congestion. Based on real-time feeds obtained
from smart-phone users, we were able to identify segments of congestion and classify them into three levels
(level $1$ represents low congestion and level $3$ represents high congestion). This is shown as a heat-map
in Figure~\ref{fig:BLR_Manyata_10_15AM_CM}, demonstrating \name{}'s effectiveness in detecting hotspots.

Another key observation, is that Figure~\ref{fig:BLR_Manyata_10_15AM_CM} shows that the region inside the
TechPark is not congested, although Figure~\ref{fig:Manyata-heat-map} shows the average speeds on these roads
to be low (inside the blue rectangle). In fact, these roads are indeed not congested, and the low speeds
are due to narrower roads with speed bumps every 20 meters. As speeds are uniformly low and there is no
significant difference between the current observed speeds and the maximum reference speed for the
road-segments, we do not classify them as congestion hotspots. This is in line with our aim of monitoring
only those regions with high variance in speeds, and not waste critical battery energy monitoring areas with
uniform speeds (we can afford to obtain samples at a lower frequency). \name{} was able to distinguish such
areas with uniformly low speeds from actual congestion
hotspots that exhibit a large deviation between the current observed speeds and the maximum speed for that
road-segment.

In addition to detecting hotspots, high sampling rate near the hotspots ensures that we obtain
sufficient samples around the hotspot region to be able to accurately characterize temporal variance at the
hotspot as shown in Figure~\ref{fig:split-time-Manyata-congestion}.

%We illustrate the effectiveness of adaptive 
%sampling through an experiment where three similar HTC Desire phones each installed one of the following 
%apps: (1) \name{} with the default adaptive sampling; (2) Modified \name{} with continuous sampling 
%where samples are taken every second (same as $S_{\operatorname{min}}$ for adaptive sampling), 
%(3) Google Maps with traffic updates~\cite{googlemaps}. 

\vspace{-0.05in}
\subsubsection{Adaptive vs. Continuous Sampling}

%\begin{figure}
%\vspace{-0.1in}
%\centering
%\includegraphics[width=80mm]{figs/AdaptiveSampling.pdf}
%\vspace{-0.15in}
%\caption{Adaptation of GPS sampling interval as user is moving away from a hotspot.}
%\label{fig:AdaptiveSampling}
%\vspace{-0.2in}
%\end{figure}

%\reminder{Move this graph to implementation after the equation for adaptive sampling.}
%We first show the behavior of adaptive sampling as the user is moving away from a congestion hotspot.
%Figure~\ref{fig:AdaptiveSampling} shows that the GPS sampling interval increases drastically and tends to
%flatten out over time as the user is moving away from a congestion hotspot.
%
%\reminder{Praveen: I do not like the definition of accuracy. Accuracy should be measured as how much time
%or how much distance elapsed within congestion zone before you detected it as so. (a) y-axis is extremely
%misleading - immediate reaction is why should time to cross congestion point depend on type of sampling}
%
%\reminder{Plot (b) requires more details of what the distance outside congestion zone is. It would be good
%to have this plot for different values of $d_{out}$}

Next, we evaluate the overhead and accuracy of \name{} with continuous and adaptive GPS sampling. For the 
continuous mode, GPS was sampled every second (same as $S_{\operatorname{min}}$ for adaptive sampling).
We consider a trip of \unit{15}{km} from TechPark to user home. Congestion zone 
is a circle with 2 km radius outside the TechPark (as shown in Figure~\ref{fig:BLR_Manyata_10_15AM_CM}).

\begin{figure*}[t]
\begin{minipage}{2.5in}
\centerline{\includegraphics[height=1.2in]{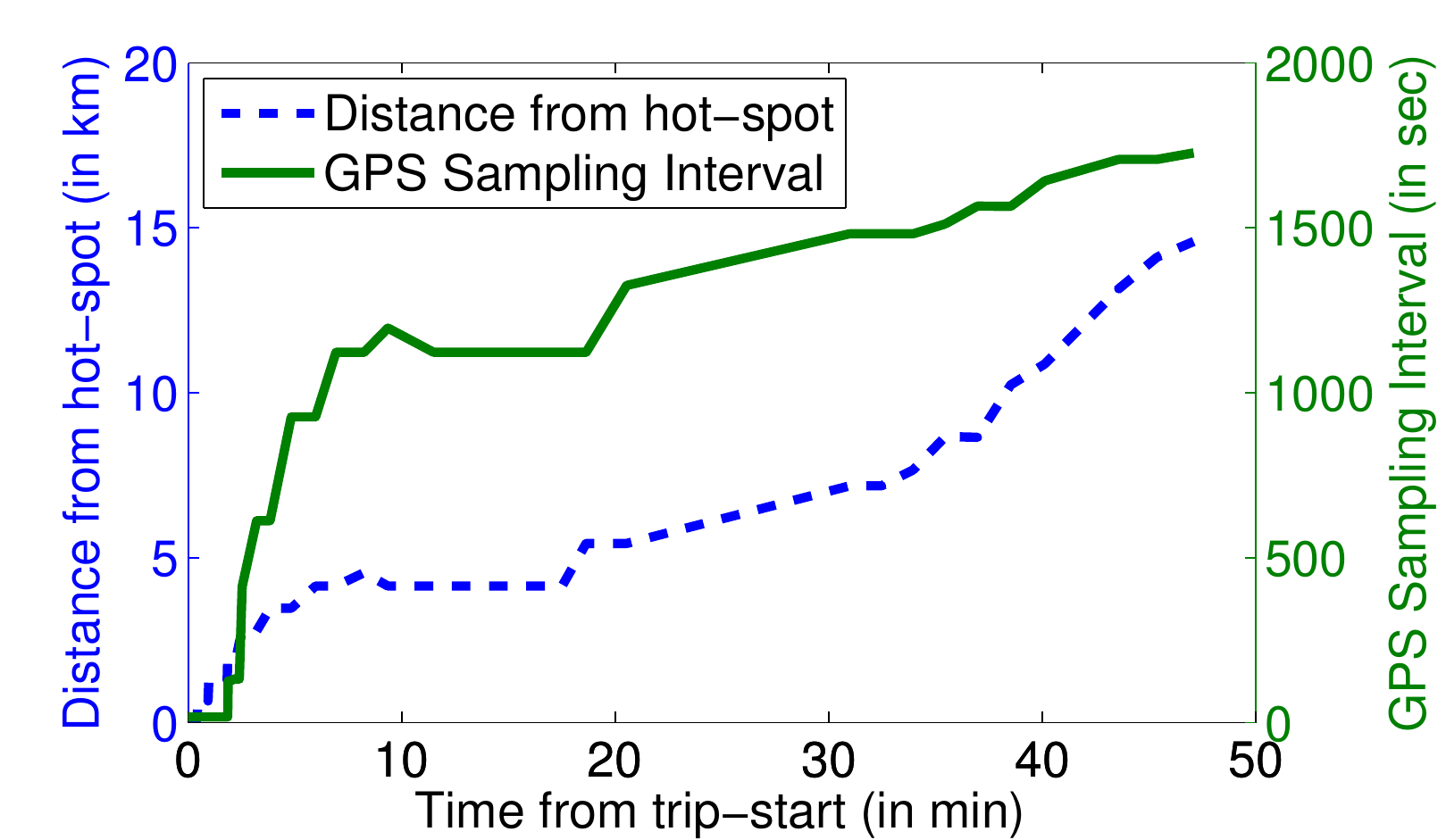}}
\caption{\label{fig:AdaptiveSampling} Adaptation of GPS sampling interval as user is moving away from a hotspot.}
\end{minipage}
~~
\begin{minipage}{1.8in}
\centerline{\includegraphics[height=1.2in]{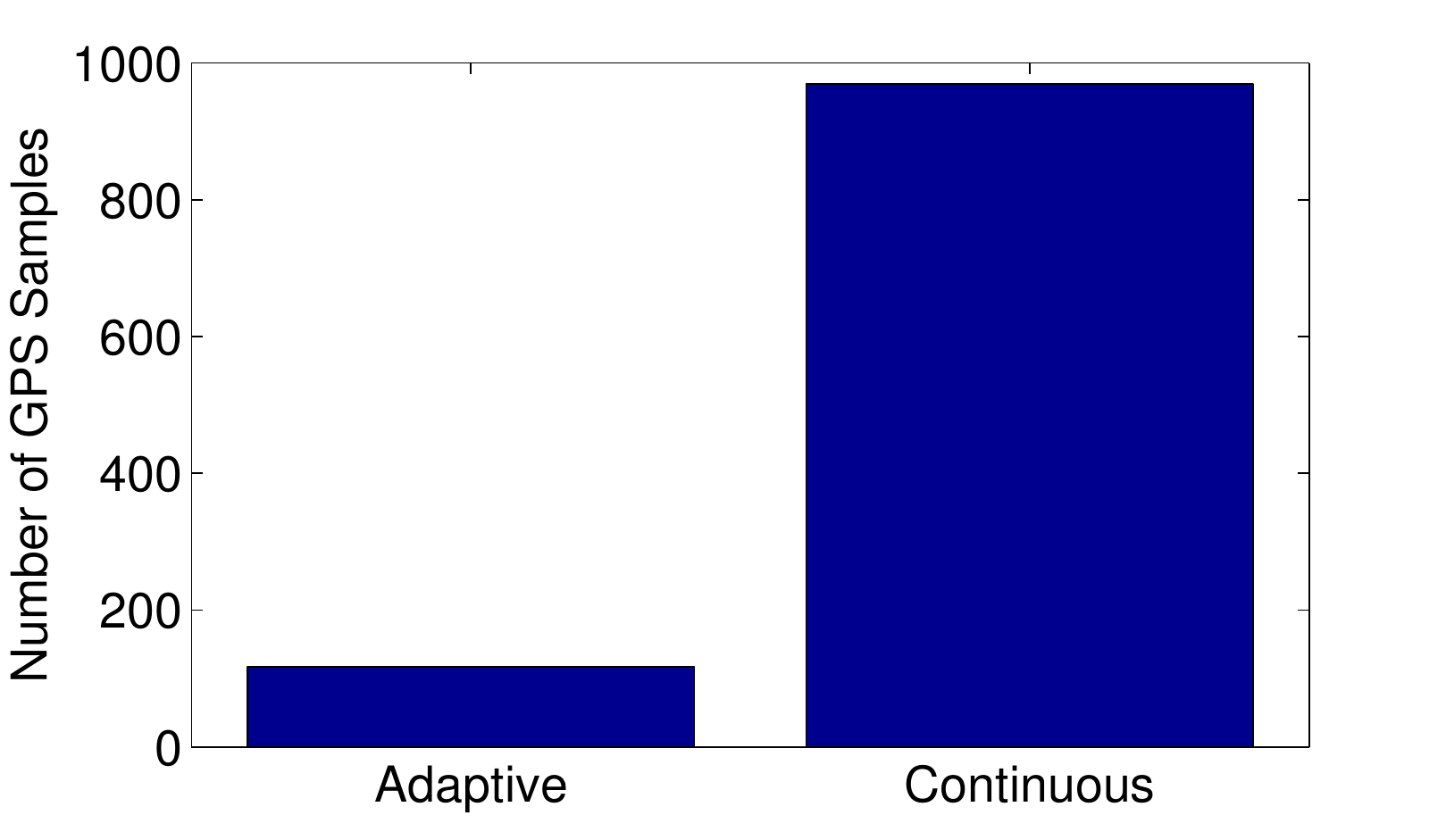}}
%\vspace{-0.2in}
\caption{\label{fig:ProbContSampling}Overhead of Continuous Vs. Adaptive Sampling}
\end{minipage}
~~
\begin{minipage}{2.5in}
\centerline{\includegraphics[height=1.2in]{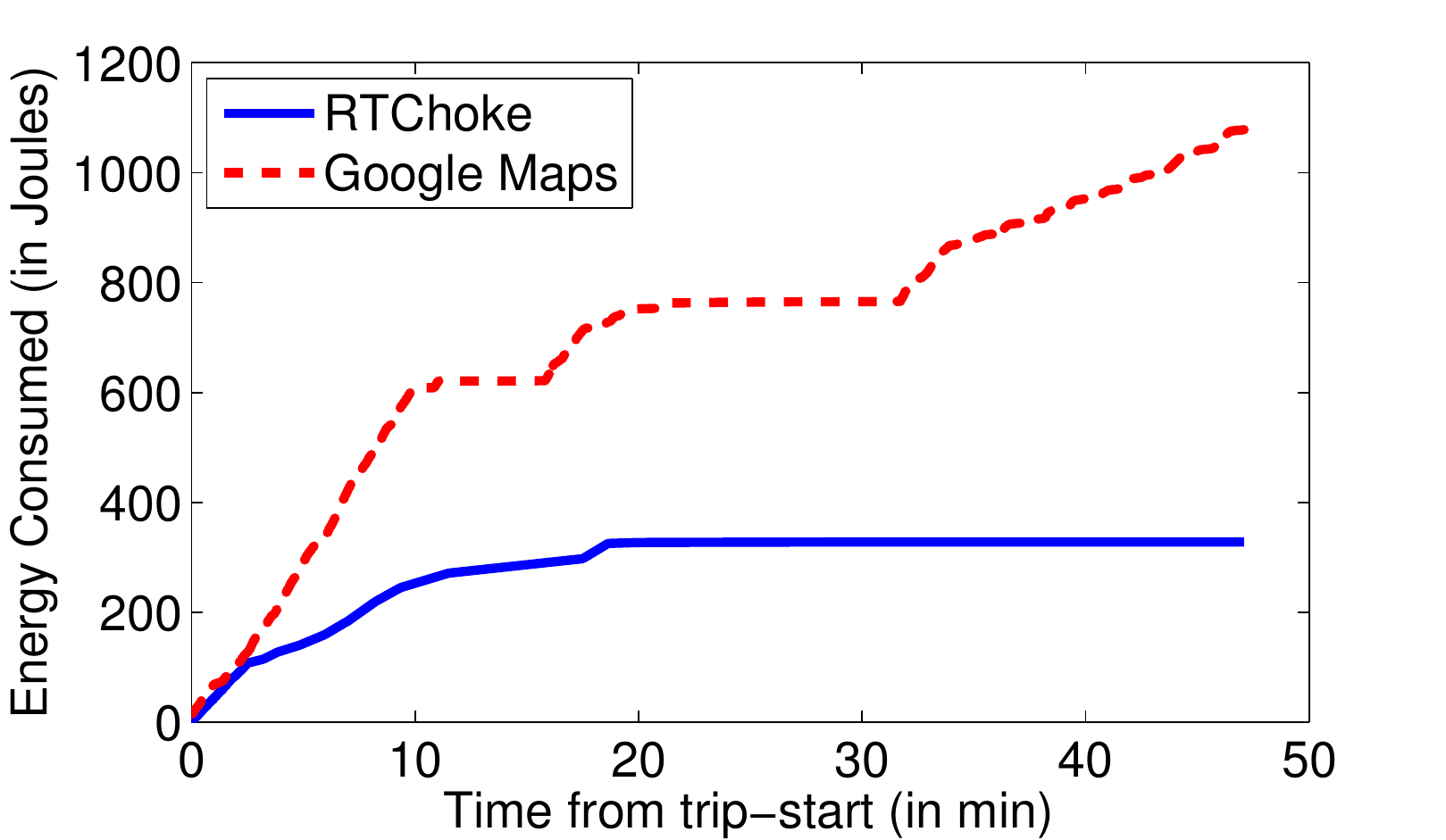}}
\caption{\label{fig:Energy_DC_Gmaps}Comparison of total energy consumed by \name{} and Google maps.}
\end{minipage}
\vspace{-0.2in}
\end{figure*}

Figure~\ref{fig:AdaptiveSampling} shows how the sampling interval increases sharply (based on 
adaptive sampling) as a typical user moved away from the hotspot zone. The sampling period is 
multiplicatively increased from $S_{\operatorname{min}}$ as the user moves away from the hotspot. 
As shown in Figure~\ref{fig:ProbContSampling}, adaptive sampling yields about $9\times$ improvement 
in terms of the overhead, captured as the number of GPS samples collected and uploaded. 

We now discuss the accuracy of adaptive sampling. We define the accuracy as the time when the user 
is detected to have entered the congestion zone when using adaptive sampling as compared to the app using 
continuous GPS sampling. We compare the time when the congestion zone was entered for 3 trips from user 
home to TechPark (reverse of the above trajectory from office to home) since it is a harder case to detect 
entrance 
We observed that the adaptive sampling took additional $45$ seconds on an average to 
detect that the user has entered the congestion zone when compared to the continuous sampling, 
which is merely $4.8\%$ of the total time spent by the user in the congestion zone.
%For a user approaching to a congestion hotspot, accuracy of the sampling mode is determined based
%on the time when the application detects that it has entered the congestion region. We consider users' trips
%during morning: from home to the office in the TechPark. Since each user may arrive at different times to the
%congestion hotspot, we track the average of time taken by users to cross the congestion hotspot.
%Figure~\ref{fig:ProbContAccuracy} shows that the accuracy of adaptive sampling is almost equal to that of
%continuous sampling. 
%% fig:AdaptiveSampling: Trip start time - 6/12/2012 22:56:19 
%% fig:Energy_DC_Gmaps: Trip start time - 6/12/2012 22:56:19 
%\begin{figure*}[t]
%\begin{minipage}{2.5in}
%\centerline{\includegraphics[width=2.5in]{figs/AdaptiveSampling.pdf}}
%\caption{\label{fig:AdaptiveSampling}Adaptation of GPS sampling interval as user is moving away from a hotspot.}
%\end{minipage}
%~~
%\begin{minipage}{1.8in}
%\centerline{\includegraphics[width=1.4in]{figs/ProbContSampling.pdf}}
%%\vspace{-0.2in}
%\caption{\label{fig:ProbContSampling}Continuous Vs. Adaptive Sampling: Overhead}
%\end{minipage}
%~~
%\begin{minipage}{2.5in}
%\centerline{\includegraphics[width=2.5in]{figs/Energy_DC_Gmaps.pdf}}
%\caption{\label{fig:Energy_DC_Gmaps}Comparison of total energy consumed by \name{} and Google maps.}
%\end{minipage}
%\vspace{-0.2in}
%\end{figure*}

%\begin{figure}
%\vspace{-0.1in}
%\begin{center}
%\mbox{
%\subfigure[
%~\label{fig:ProbContAccuracy}]{\includegraphics[scale=0.3]
%{figs/ProbContAccuracy.pdf}}
%%\quad
%\subfigure[
%~\label{fig:ProbContSampling}]{\includegraphics[scale=0.3]
%{figs/ProbContSampling.pdf}}}
%\end{center}
%\vspace{-0.15in}
%\caption{Continuous Vs. Adaptive Sampling: \subref{fig:ProbContSampling} Accuracy; \subref{fig:ProbContSampling} Overhead.}
%\vspace{-0.2in}
%\end{figure}

\vspace{-0.05in}
\subsubsection{Energy Footprint}

We compare the energy footprint of \name{} with Google Maps~\cite{googlemaps} app with real-time traffic updates enabled. To estimate application-specific energy consumption, we use PowerTutor~\cite{powertutor},  which provides per-application energy consumption by different components, such as CPU, 3G, and LCD screen. 
Google maps with traffic updates enabled, can be used to view the status of congestion on different roads. Unlike the adaptive sampling of \name{}, Google maps samples GPS continuously and uploads location data in real-time to the server. It retrieves traffic information in terms of low-, medium-, and high-speeds from their server(s). The only additional operation performed by Google maps is retrieving map-specific data. \name{} does not require map-specific
information for its functioning. For fairness of comparison, we perform the experiment by running \name{} and Google maps on two identical HTC Desire smartphones, both carried by the same user. Also, we subtract the LCD screen specific energy consumption from the total energy to measure only the energy consumed by the application.

%\begin{figure}
%\vspace{-0.1in}
%\centering
%\includegraphics[width=80mm]{figs/Energy_DC_Gmaps.pdf}
%\vspace{-0.15in}
%\caption{Comparison of total energy consumed by \name{} and Google maps.}
%\label{fig:Energy_DC_Gmaps}
%\vspace{-0.1in}
%\end{figure}

Figure~\ref{fig:Energy_DC_Gmaps} compares energy consumed by \name{} and Google maps over time, when a user
(carrying both the phones) is moving away from a congestion hotspot. Note that for the first $170$ seconds,
the user is within the congestion zone, where \name{} samples and uploads at the peak rate. Thus, \name{}
consumes almost as much energy as Google maps. However, Google map's energy consumption increases
almost linearly even beyond the congestion region, but energy for \name{} grows much slower
because of adaptive sampling.

We also measured the total uplink traffic from \name{} and Google maps (graph omitted in the interest of space).
The total bytes uploaded by Google maps was $85.5$ KB, whereas \name{} uploaded $19.36$ KB data, a
reduction in network usage by about $4.5$ times.

\begin{figure}
\centering
\includegraphics[width=85mm,height=35mm]{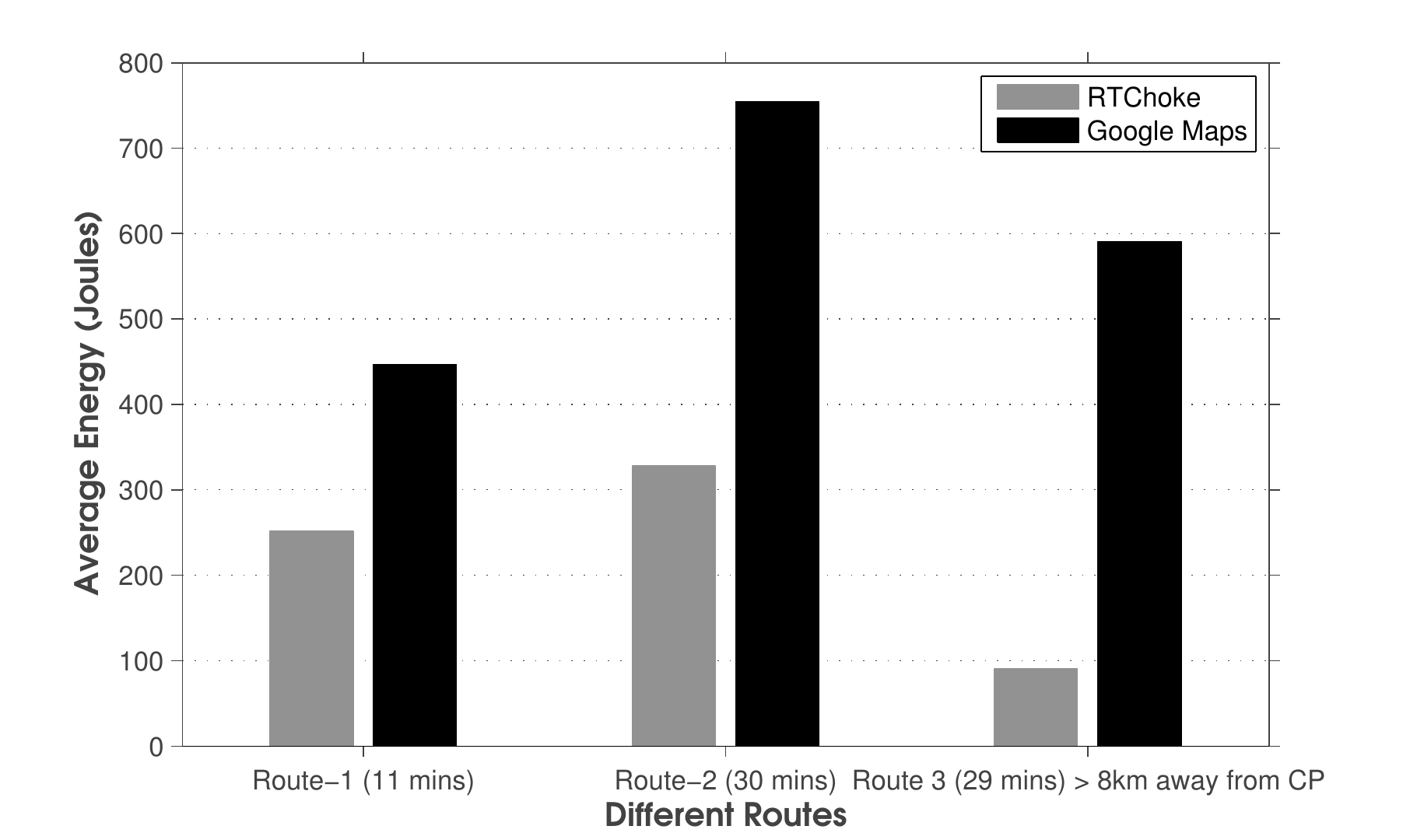}
\vspace{-0.1in}
\caption{Energy consumption on \name{} and Google Maps.}
\label{fig:avgEnergy}
\vspace{-0.2in}
\end{figure}

To avoid route-specific biases, we also compared the average energy consumption for three different routes
with different travel durations as shown in Figure~\ref{fig:avgEnergy}. It can be observed that the difference
in energy consumption between Google maps and \name{} is higher when the duration of travel away from a
congestion hotspot is higher, as demonstrated by Route-1 and Route-2. When the user is
traveling a longer distance outside the congestion hotspot (8 km away for Route-3), adaptive sampling
yields an order of magnitude lower energy footprint compared to that of Google maps.

%For time between$600$ and $1000$ seconds, Google map's energy consumption seems to be constant. We are
%unable to explain this behavior--it could be %because of some optimizations incorporated in Google maps.

%% file: experiments_sfoTR.tex
\vspace{-0.1in}
\subsection{Congestion in San Francisco}
%\begin{figure}
%\centering
%\includegraphics[width=75mm]{figs/SFO_500Cabs_9AM_CM.eps}
%\caption{Congestion hotspots in San Francisco.\reminder{This needs to be removed}}
%\label{fig:SFO_500Cabs_9AM_CM}
%\end{figure}

We analyzed location information from 200 cabs across 3300 road segments in San Francisco city using the 
CRAWDAD data~\cite{comsnets09piorkowski}. Each cab reports GPS coordinates and timestamp once every minute. 
We process this data as explained in Section~\ref{sec:design} to estimate the congestion in SFO city. 

%\begin{figure*}
%\begin{center}
%\mbox{
%\subfigure[CDF of speeds observed on SFO roads.
%~\label{fig:avgSpeedCDF}]{\includegraphics[scale=0.5]
%{figs/avgSpeedCDF.eps}}
%\quad \subfigure[Congestion hotspots vary during different time-of-the-day.
%~\label{fig:hotSpotsPerTime}]{\includegraphics[scale=0.5]
%{figs/hotSpotsPerTime.eps}}
%        }
%\end{center}
%\caption{Speed analysis in SFO}
%\end{figure*}

%\begin{figure*}
%\vspace{-0.1in}
%\begin{center}
%\mbox{
%\subfigure[Hotspots vary with time-of-day.
%~\label{fig:hotSpotsPerTime}]{\includegraphics[width=5cm]
%{figs/hotSpotsPerTime.pdf}}
%\subfigure[Peak hour (17:00 PM) heat-map
%~\label{fig:maxCM_SFO}]{\includegraphics[width=5cm]
%{figs/maxCM_SFO.pdf}}
%\quad \subfigure[Off-peak hour (04:30 AM) heat-map.
%~\label{fig:minCM_SFO}]{\includegraphics[width=5cm]
%{figs/minCM_SFO.pdf}}
        %}
%\end{center}
%\vspace{-0.2in}
%\caption{Temporal variation and heat-maps of hotspots in SFO}
%\label{ref:sfoHeat}
%\vspace{-0.1in}
%\end{figure*}

\begin{figure}
\centering
\includegraphics[scale=0.4]{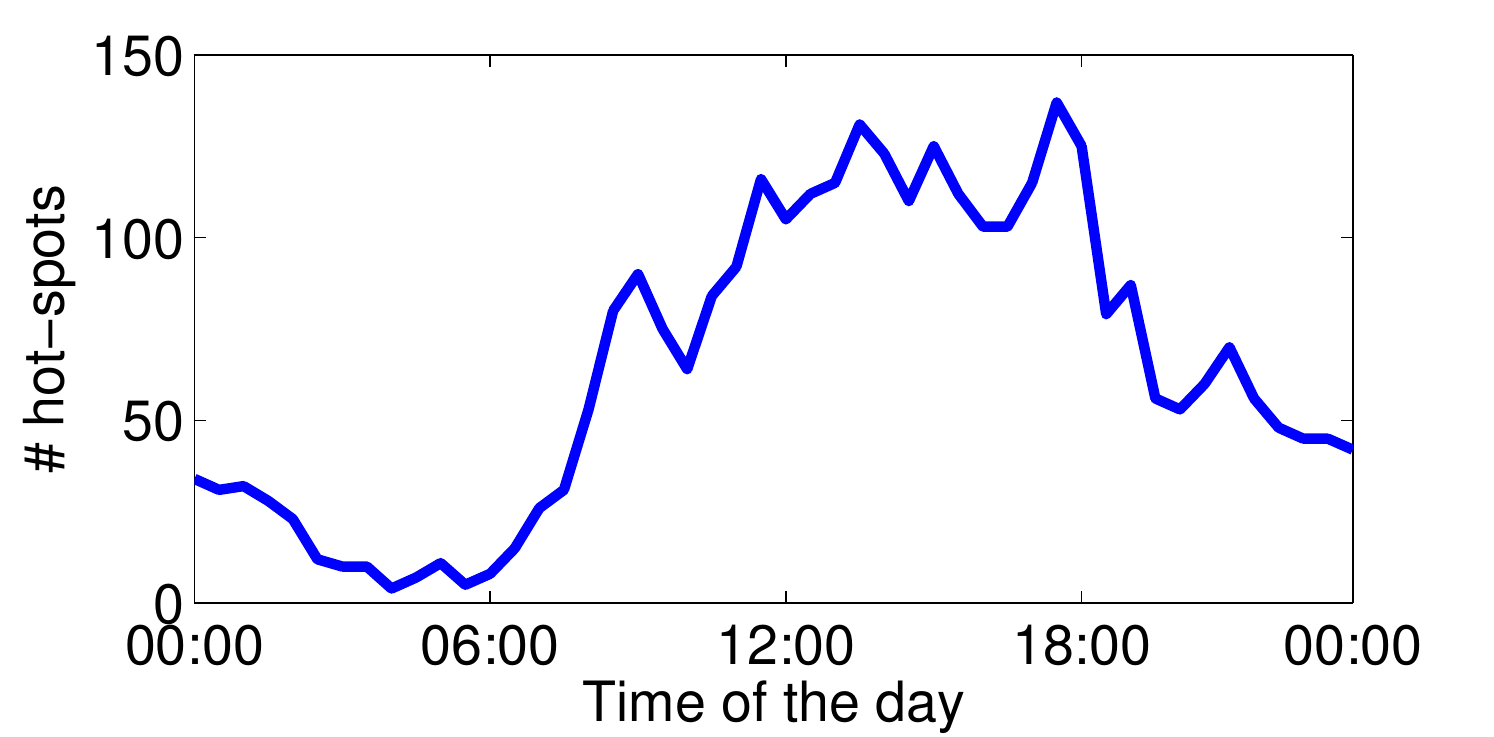}
\caption{SFO: Hotspots vary with time-of-day}
\label{fig:hotSpotsPerTime}
\end{figure}
\begin{figure}
\centering
\includegraphics[scale=0.4]{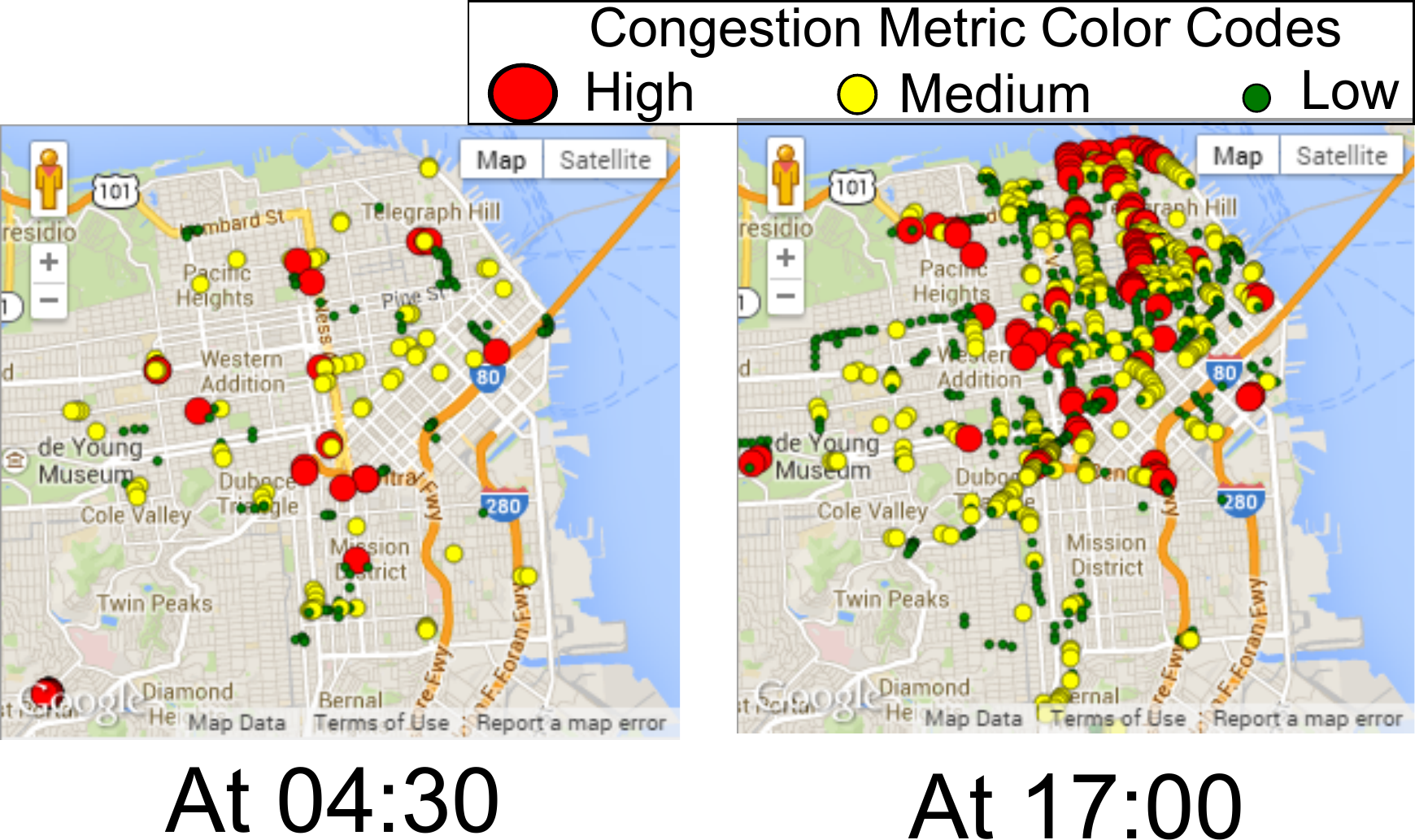}
\caption{Off-peak and Peak hour congestion heat-map}
\label{fig:sfo_peaks}
\end{figure}

\begin{figure*}
\centering
\includegraphics[scale=0.4]{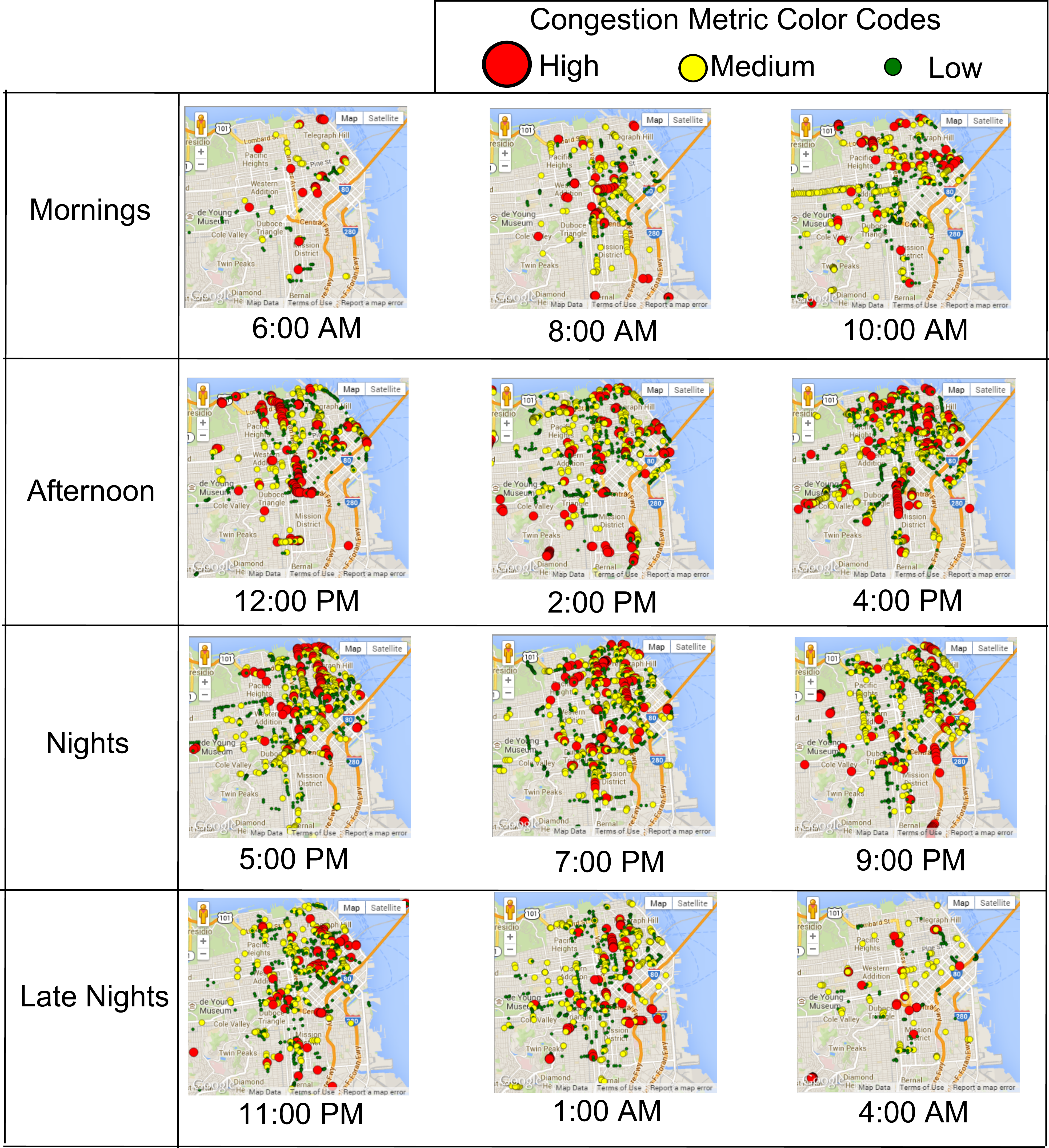}
\caption{Spatio-temporal variation of congestion hot-spots in SFO}
\label{fig:sfo_st}
\end{figure*}

We use the temporal speed data per road-segment to compute the congestion levels in SFO city. The median 
average speed on road-segments is around \unit{45}{kmph} (or \unit{28}{mph}). Fig.~\ref{fig:hotSpotsPerTime} 
shows the temporal variance of hotspots in SFO city; the number of hotspots vary with the time-of-day 
showing peak and off-peak traffic. Fig.~\ref{fig:sfo_peaks} shows the congestion heat-map at peak 
traffic (at 17:00), with hotspots marked in red. Only 137 roads (approximately 4\% of the all roads 
sampled) were tagged as hotspots, where users should expect large delays due to lower speeds. During the 
off-peak hours (at 04:30) the number of congested road-segments was even smaller. Fig.~\ref{fig:sfo_st} shows the spatio-temporal variations of SFO congestion 
hot-spots at different times of the day.

%\begin{figure}
%\vspace{-0.1in}
%\centering
%\includegraphics[scale=0.3]{figs/devAvgRatioHotSpot.eps}
%\vspace{-0.15in}
%\caption{Deviation in speed at hotspots are larger than at non hotspots}
%\label{fig:devAvgRatioHotSpot}
%\vspace{-0.1in}
%\end{figure}

\begin{figure*}[t]
\begin{minipage}{2.2in}
\centerline{\includegraphics[width=2.2in]{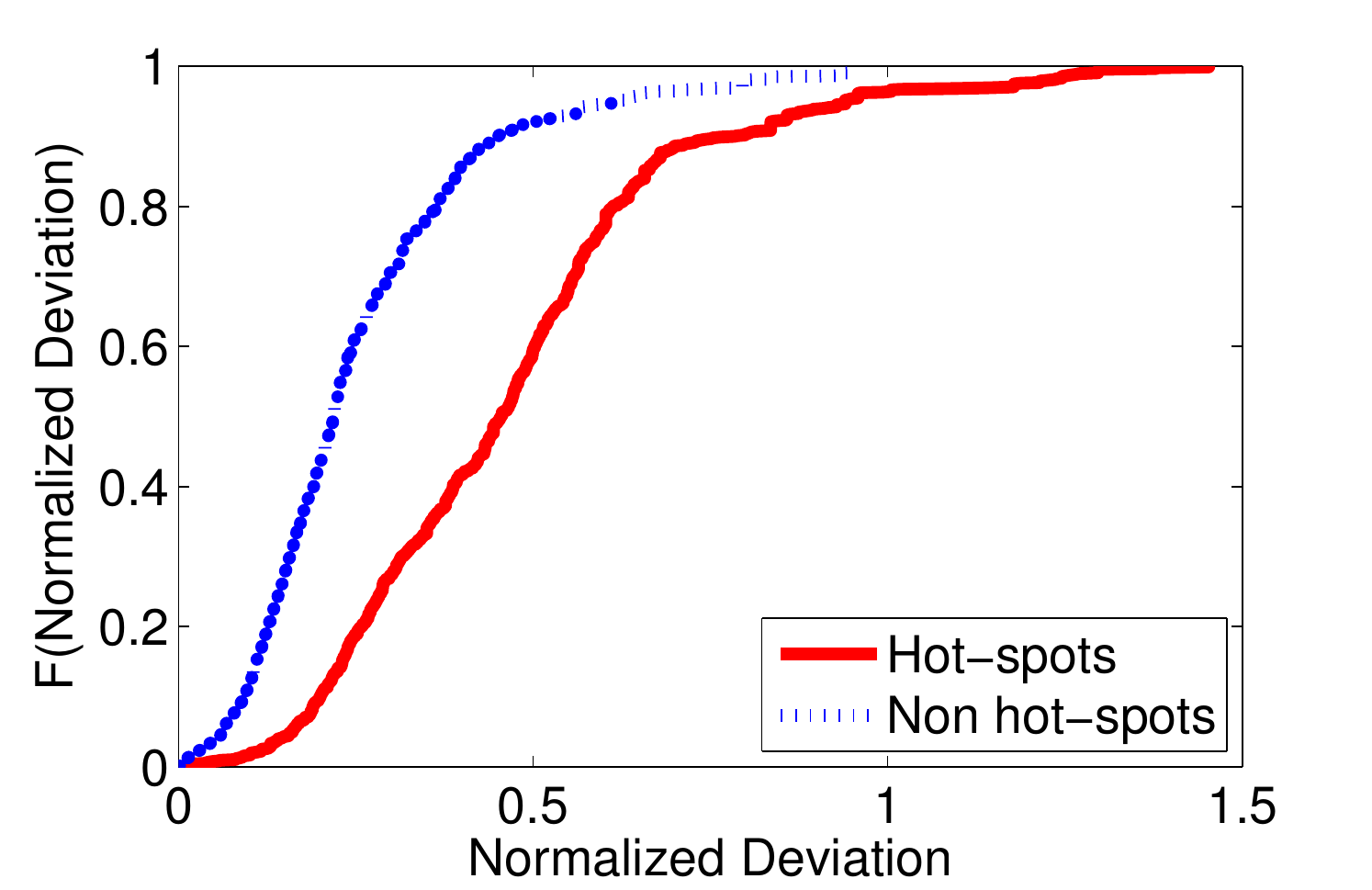}}
\caption{\label{fig:devAvgRatioHotSpot}Deviation observed on hotspots are larger than those observed on non hotspots}
\end{minipage}
~~
\begin{minipage}{2.3in}
\centerline{\includegraphics[width=2.3in]{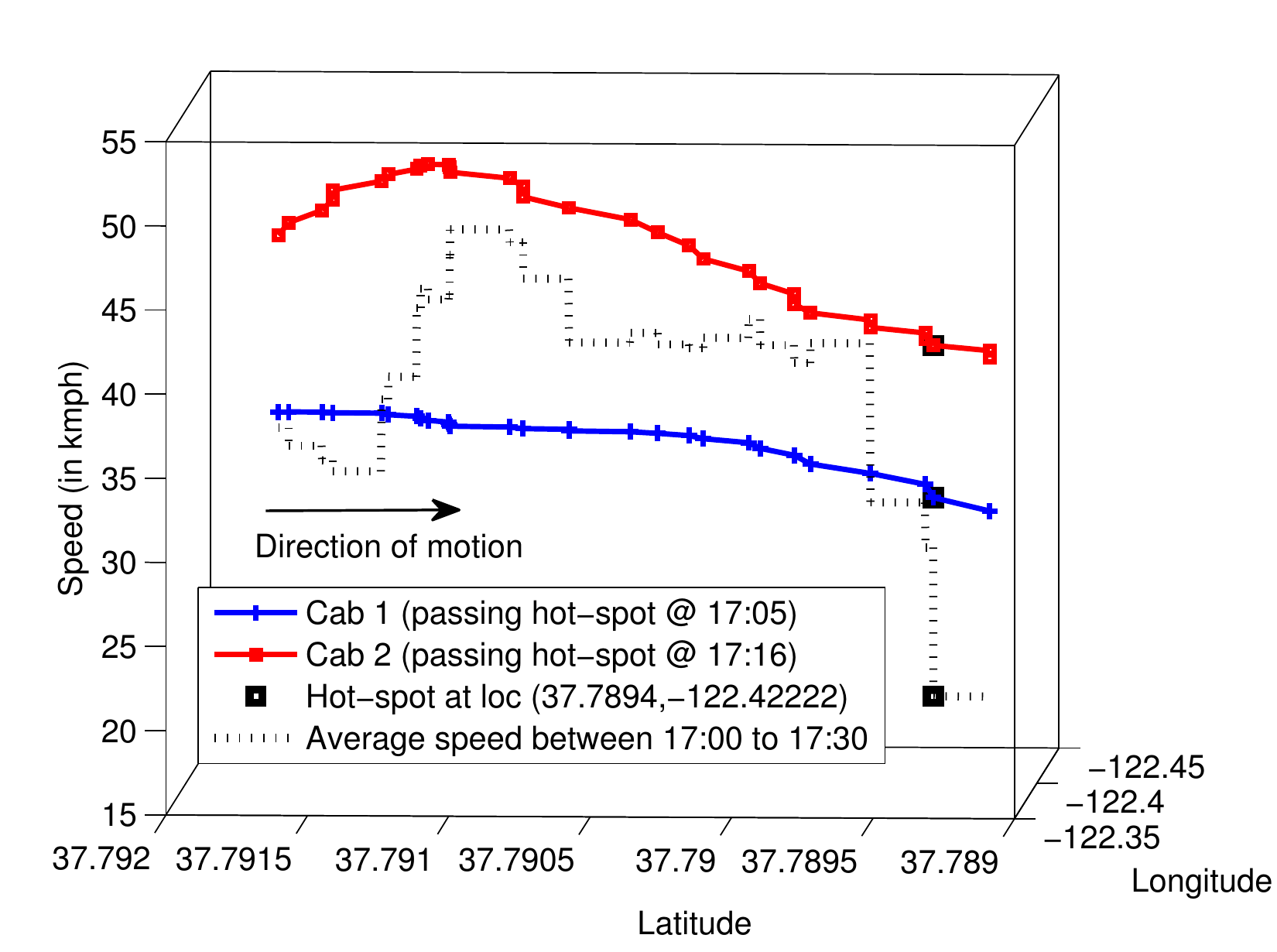}}
%\vspace{-0.2in}
\caption{\label{fig:smoothedSpeedsTraj3D}Two trajectories intersecting at a hotspots}
\end{minipage}
~~
\begin{minipage}{2.3in}
\centerline{\includegraphics[width=2.3in,height=1.8in]{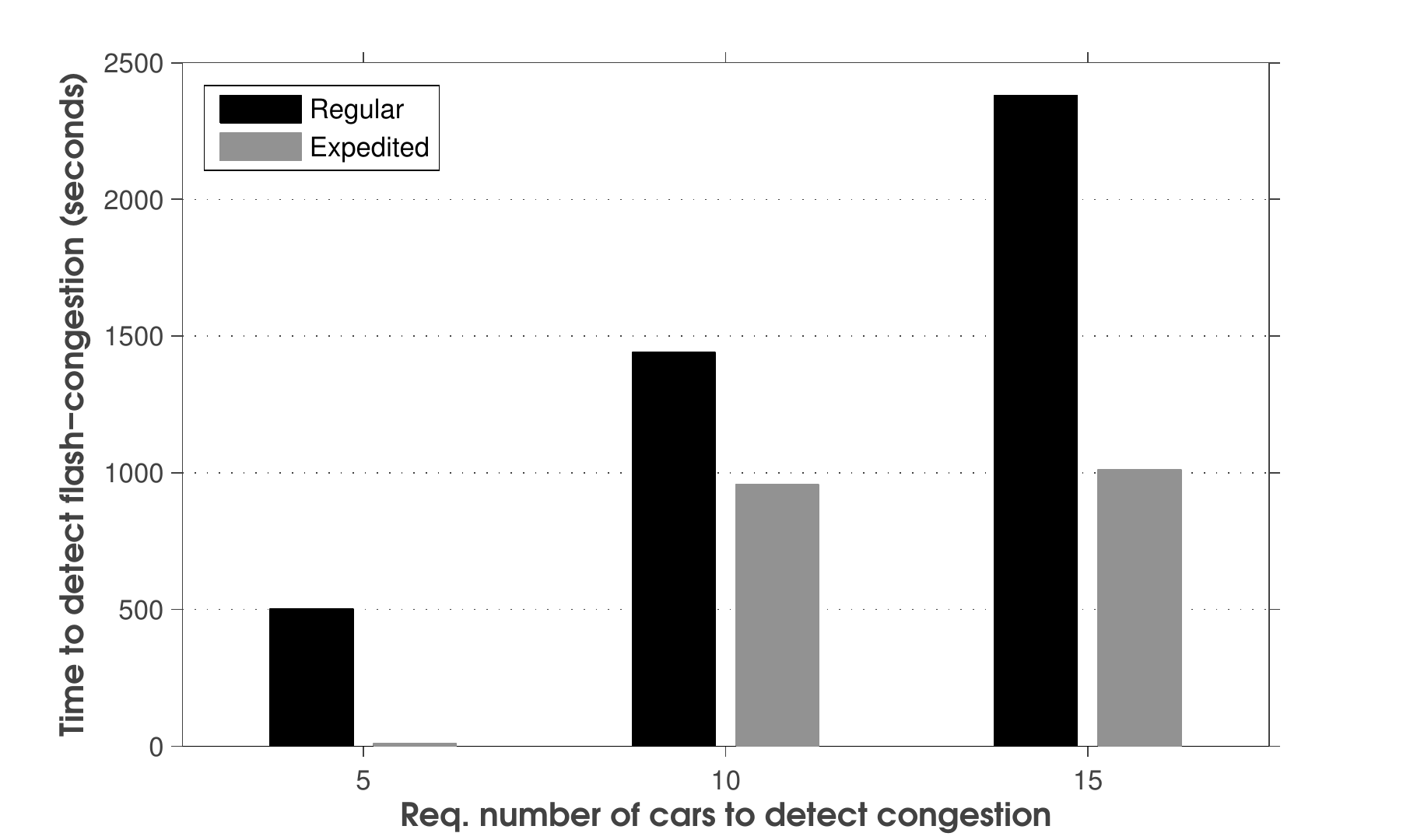}}
\caption{\label{fig:FlashCongDetTime}Time to detect flash congestion.}
\end{minipage}
\vspace{-0.1in}
\end{figure*}

Similar to TechPark data, we observed that hotspots demonstrated a much larger variation in speeds when 
compared to the other road-segments. In order to measure how the road-speeds vary in congestion hotspots, 
we compute the Normalized Deviation (ND) of each road-segment, defined as the ratio of the standard 
deviation to the average speed on the road-segment. 
%Higher values of ND suggest large deviation relative to the average speed. 
Fig.~\ref{fig:devAvgRatioHotSpot} plots the CDF of ND for hotspots and 
non hotspots. The figure shows that hotspots have larger values of ND, and hence larger temporal variation. 

%\begin{figure}
%\vspace{-0.05in}
%\centering
%\includegraphics[scale=0.3]{figs/smoothedSpeedsTraj3D.eps}
%\vspace{-0.15in}
%\caption{Two trajectories intersecting at a hotspot}
%\label{fig:smoothedSpeedsTraj3D}
%\vspace{-0.15in}
%\end{figure}

Fig.~\ref{fig:smoothedSpeedsTraj3D} shows the trajectories followed by two cabs at 
17:05 and 17:16 on the same day. It also shows the average speed of the roads between 
17:00 to 17:30. A square marks the congestion hotspot. Note that the cabs were
traveling much faster outside the hotspot than closer to it. 
%The decrease of speeds of 
%two cabs and the large variation near the hotspot is seen as the vehicles are approaching the hotspots.

We recommend actively monitoring a small number of hotspots instead of all road segments based on our
observation that deviation in speeds is higher at hotspots and lower at non-hotspots. For the non hotspots, 
average speeds at the given time-bin over all history can be used as an approximate speed. Such selective 
monitoring of hotspots with high frequency can be used for low-complexity and energy-aware congestion 
monitoring for applications (e.g., alerting users about possible congestion and
suggesting alternate routes). Adaptive Sampling approach suggested in Section~\ref{sec:adaptiveSampling} 
utilizes this insight to monitor frequently at the hotspots, and to save energy by low-rate monitoring at 
other road-segments.

{\bf Flash congestion detection:}
We now compare the time required to detect flash-congestion. We simulate flash-congestion caused by an 
accident, in which all the cars approaching the accident location get blocked. We use the real location traces 
for cars from the San Francisco cabs data set. We assume that multiple cars (as varied on X-axis) need to 
report low speeds to detect congestion reliably. We consider $S_{\operatorname{min}} = 1$ minute (the minimum granularity of
samples in the data set) and $S_{\operatorname{max}} = 2$ minutes. Since there was no congestion before, we use $S_{\operatorname{max}}$ as 
the cars' sampling interval. We consider two types of alerting modes: {\em Regular} and {\em Expedited}. 
In the {\em Regular} mode, cars do not change their sampling interval until congestion is detected reliably. 
On the other hand, in the {\em Expedited} mode, the server declares that a reported location is a 
{\em possible} congestion hotspot, if even a single car notifies lower speed than threshold. In such a case, 
the server notifies the other cars approaching the same {\em possible} congestion point to sample at peak 
rate (at $S_{\operatorname{min}}$). 

Fig.~\ref{fig:FlashCongDetTime} plots the time required to reliably detect 
flash-congestion after the first car has notified a lower-speed value than the congestion threshold. It can be 
seen that the time to detect flash congestion can be reduced by up to $50\times$ using {\em Expedited} alerting 
mode. The significant increase in congestion detection time from that for $5$ versus $10$ or $15$ cars is due 
to the cabs arrival rate in the dataset.

%% file: conclusion.tex
\section{Conclusion}
\label{sec:conclusion}

In this paper, we show that in a traffic monitoring system built over mobile sensors, automatically detecting and monitoring congestion hotspots can significantly reduce the overall overhead of the system, without compromising on the accuracy, in comparison, to a system with fine-grained periodic monitoring.
The solution, \name, requires minimal user intervention, and optimizes both battery power and network bytes used for data uploads. We believe that detecting hotspots and monitoring congestion just based on hotspots makes participatory sensing traffic applications efficient by coupling collection and use of data.
% however, other instantiations of hotspots are possible, which can form interesting future work.

%Through this work, we demonstrate that collecting data based on its utility (i.e. having a feedback loop between the sensors uploading information and the servers using this information) systematically minimizes the overhead of such participatory sensing systems.